\documentclass[usenatbib,usegraphicx]{mn2e}
\usepackage{amsmath}
\usepackage{times}
\usepackage{subfigure}

\newcommand{\fmag}{\hbox{$.\!\!^m$}}

\newcommand{\Rj}{\hbox{R$_{jup}$\,}}

\newcommand{\Rsol}{\hbox{R$_{\odot}$\,}}

\begin{document} 
\title[Searching for Planetary Transits in the Field of Open Cluster NGC~6819]
{Searching for Planetary Transits in the Field of Open Cluster NGC~6819 - I.
\thanks{Based on observations made with the 
Isaac Newton Telescope operated on the island of La Palma by the Isaac Newton 
Group in the Spanish Observatorio del Roque de los Muchachos of the Instituto 
de Astrofisica de Canarias.}\\ }
\author [R.~A.~Street et al.]{\parbox[t]{\textwidth}{R.~A.~Street$^{1,2}$,
Keith Horne$^1$, T.~A.~Lister$^1$, A.~J.~Penny$^3$, Y.~Tsapras$^{1,4}$, 
A.~Quirrenbach$^5$, N.~Safizadeh$^5$, D.~Mitchell$^5$, 
J.~Cooke$^5$, A.~Collier~Cameron$^1$} \\
$^1$ School of Physics and Astronomy, University
of St.~Andrews, North Haugh, St. Andrews, Fife, KY16 9SS, Scotland \\ 
$^2$ APS Division, School of Mathematics \& Physics, Queen's University of Belfast, 
University Road, Belfast, BT7 1NN, Northern \\ Ireland \\
$^3$ Rutherford Appleton Laboratory, Chilton, Didcot, Oxon, OX11 0QX,
England\\
$^4$ School of Mathematical Sciences, Queen Mary University of London, 
Mile End Road, London, E1 4NS, England \\
$^5$ Center for Astrophysics and Space Sciences (CASS), 
University of California, San Diego, 9500 Gilman Drive, La Jolla, \\
CA 92093-0424, USA \\
{\rm(email: R.Street\@@qub.ac.uk)}}
\maketitle

\begin{abstract}

We present results from our survey for planetary transits in the field of the
intermediate age ($\sim$2.5\,Gyr), metal-rich ([Fe/H]$\sim$+0.07) open cluster
NGC~6819.  We have obtained high-precision time-series photometry for over
38,000 stars in this field and have developed an effective matched-filter
algorithm to search for photometric transits.   This algorithm identified 8
candidate stars showing multiple transit-like events, plus 3 stars with single
eclipses.  On closer inspection, while most are shown to be low mass stellar
binaries, some of these events could be due to brown dwarf companions.  The data
for one of the single-transit candidates indicates a minimum radius for the
companion similar to that of HD~209458b.

\end{abstract}

\begin{keywords}
methods: data analysis -- open clusters and associations: general -- open
clusters and associations: individual: NGC~6819 -- binaries: eclipsing -- stars:
low mass, brown dwarfs -- planetary systems.
\end{keywords}

\section{Introduction}

Following the discoveries of Hot Jupiter-type extra-solar planets, it was
apparent that a photometric survey to detect these planets in transit was
feasible from ground-based telescopes.  This was confirmed by the detection of
transits by HD~209458b by \citet{charbonneau2000} and \citet{henry2000}.  Such
a survey has the potential to broaden the search for planets over a greater
volume of space by monitoring stars at fainter magnitudes than are accessible
to the spectroscopic radial velocity technique.  The transit method enables us
to probe far greater numbers of stars {\em simultaneously} and so obtain a
statistically significant sample of planets in a comparatively short time. 
Discoveries can be used to determine the abundance of planets in a range of
stellar environments, allowing us to investigate the relationship between
planet formation and key properties; for example metallicity, age, stellar and
radiation density.  Follow-up observations of transiting planets can provide
vital information on individual planets -- their true masses, radii and orbital
inclinations -- crucial for testing theories of planetary structure.  Once
transit candidates are obtained from photometric surveys, these data combined
with spectrographic observations will not only provide this information but
will also distinguish planetary transits from grazing-incidence eclipses by
stellar and brown dwarf companions.  Follow-up observations are discussed in
greater detail in Section~\ref{sec:followup}.  

From the radial velocity survey results, we know that approximately 1 percent
of F--K-type solar neighbourhood stars harbour hot Jupiters (giant planets with
periods of $\sim$2--6\,days).  The typical orbital separation (0.05\,AU) for
these planets implies that about 10\% of them should exhibit transits if their
orbital planes are randomly oriented with respect to the line of sight. 
Assuming that planetary orbits are distributed isotropically, we can expect
there to be one transit for every 1000 stars.  The detection probability is
affected by factors such as the sampling rate of the observations, the
efficiency of detection from the data and the fraction of late-type dwarfs in
the sample.  Correspondingly, transit surveys that monitor tens of thousands of
stars simultaneously may expect to detect tens of planets.  This requirement
must then be balanced against the need to avoid fields so crowded that blending
makes stars difficult to measure precisely.

\citet{janes96} suggested that open clusters would make a good compromise for
ground-based surveys.  They provide large numbers of stars within a relatively
small field with minimal blending.  They also provide a distinct population of
stars of known age and metallicity and allow us to probe stars in the
environment where they formed.  Additionally, these fields provide a separate
population of background stars for comparison.  

A number of groups are pursuing transit surveys using large (2--4m) telescopes
with wide field, mosaic CCD cameras.  Most notably, the {\sc explore} project
has used the CTIO-4m and the 3.6m CFHT to observe two Galactic plane fields in
2001 \citep{yee02}.  These data have revealed 3 possible planetary transit
candidates, and the team are in the process of obtaining radial velocity
follow-up \citep{mallen02}.  The {\sc ogle} group have used their microlensing
observations of Galactic disc stars to search for transits in the lightcurves
of $\sim$52,000 stars, yielding 59 candidates so far \citep{udalski02}. 
\citet{dreizler02} obtained classification spectra for 16 of these stars,
allowing them to estimate the radius of the primary and infer the radius of the
companion.  This analysis ruled out 14 candidates as having stellar-mass
companions, while the companions of two objects were found to have radii
similar to that of HD~209458b.  Recently, \cite{mochejska02} have undertaken a
survey of open clusters using the F.~L. Whipple Observatory's 1.2m telescope,
discovering 47 new, low amplitude variables.  

In 1999 we began a survey of three open clusters for planetary transits:  
NGC~6819, 7789 and 6940.  The wide field of view of the Isaac Newton
Telescope's (INT) Wide Field Camera (WFC) is ideally suited to this task,
and we were awarded a total of 3 bright runs of 10 nights each in
June/July 1999 and September 2000 for these observations.

Open cluster NGC~6819 was observed during the first 19 of these nights and the
results presented here stem from our analysis of these data.  The basic
parameters of this cluster are given in Table~\ref{tab:N6819data}.  Previous
relevant work on this cluster was discussed in \citet{street02}; most important
is a recent study by \citet{kalirai01} which provided $B$ and $V$ magnitudes
for large numbers of stars in this field.  

\begin{table} 
\centering 
\caption{Basic data on NGC 6819.  Data taken from Kalirai et al. (2001) and 
the {\sc simbad} database.}
\protect\label{tab:N6819data} 
\vspace{5mm} 
\begin{tabular}{cc} 
\hline 
RA (J2000.0)  & 19$^{\rmn{h}}$ 41$^{\rmn{m}}$ \\ 
Dec (J2000.0) & +40$\degr$11\arcmin  \\ 
$l$           & 73$\degr$.97      \\ 
$b$           & +8$\degr$.48      \\
Distance (pc) & 2754 $\pm$ 305    \\
Radius        & $\sim$9.5\arcmin \\
Age (Gyr)     & 2.5               \\
$[Fe/H]$      & +0.07             \\ 
$E(B-V)$      & 0.10              \\
\hline 
\end{tabular} 
\end{table}

In the rest of this paper, Section~\ref{sec:obs} details these observations,
and Sections~\ref{sec:red} and \ref{sec:colours} present the data reduction
procedure.  Section~\ref{sec:tf} discusses our transit-detection algorithm
while our results are presented in Section~\ref{sec:results}.  Finally we draw
conclusions in Section~\ref{sec:conc}.  

\section{Observations}
\protect\label{sec:obs}

We observed NGC 6819 on 19 nights during 1999 June 22--30 and 1999 July
22--31, using the 2.5m Isaac Newton Telescope, La Palma.  The Wide Field
Camera (WFC) employs four 2048$\times$4096 pixel EEV CCDs to image a
$\sim$0.5$\degr\times$0.5$\degr$ field of view with a pixel scale of
0.33$\arcsec$\,pix$^{-1}$.  The gain and readout noise values for each CCD
were taken from the Cambridge Astronomical Survey Unit
webpage\footnote{http://www.ast.cam.ac.uk/$\sim$wfcsur/ccd.html} and are
listed in Table~\ref{tab:CCDvalues}.

\begin{table} 
\centering 
\caption{Gain and readout noise values for the four WFC CCDs, taken from the 
Cambridge Astronomical Survey Unit webpage.}
\protect\label{tab:CCDvalues} 
\vspace{5mm} 
\begin{tabular}{ccc} 
\hline 
CCD No.		      	     &	Gain (e$^{-}$\,ADU$^{-1}$) & Readout noise (ADU) \\
\hline
1			     &       3.12		   & 7.9  \\
2			     &       3.19		   & 6.4  \\
3			     &       2.96		   & 8.3  \\
4			     &       2.22		   & 8.3  \\
\hline
\end{tabular} 
\end{table}

\begin{figure*}
\begin{tabular}{c}
\includegraphics[angle=0,width=8cm]{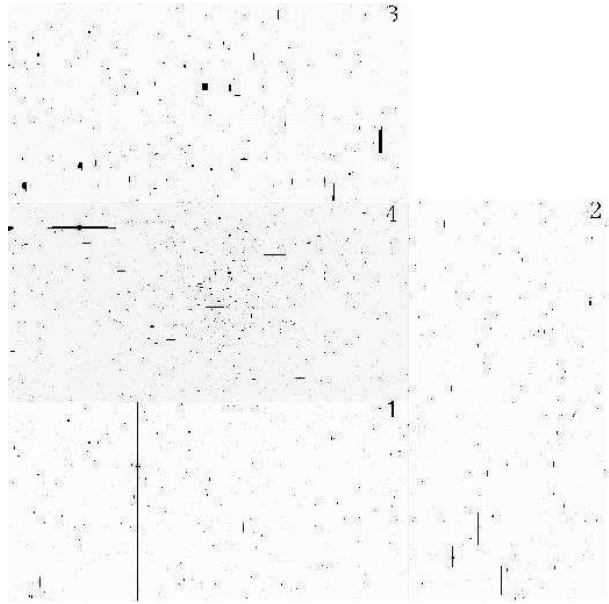}
\end{tabular}
\caption{The WFC $\sim$0.5$^{\degr}\times$0.5$^{\degr}$ field of view covering 
open cluster NGC~6819; North is to the top of the image while East is to the 
left.  The CCD numbers are given in the corners.}
\protect\label{fig:cluster}
\end{figure*}

Three open clusters, NGC 6819, 6940 and 7789, were observed in rotation during
these two runs, taking pairs of 300s exposures through the Sloan r' filter on
each visit.  The readout time of the WFC at the time required 160\,sec of
deadtime between exposures.  No dithering was applied between exposures as we
aimed to place each star as close to the same pixel each time as possible.  In
practise, the $x,y$ shifts between images were up to a few pixels.  NGC~6819
was observed for $\sim7-8$ hours each night, typically resulting in 16 -- 25
frames per night or about 2 frames/hour.   In total, 361, 384, 364 and 325
frames were obtained of this cluster for CCDs 1 -- 4 respectively.  The number
of available frames varied due to unpredictable readout failures which could
affect individual CCDs.  The average gap between pairs of exposures was, at
most, roughly an hour and we had good observing conditions on all nights.  The
field of view covered by the WFC is shown in Figure~\ref{fig:cluster}.

\section{Data Reduction}
\protect\label{sec:red}

We developed a semi-automated data reduction pipeline in order to process this
large dataset.  This pipeline has been previously described in \citet{street02}
and in detail in \citet{streetth}.  

The debiasing, flatfielding and trimming of the frames were carried out using
the Starlink package {\sc figaro} \cite{figaro}.  A correction for the
nonlinear response of each chip was also applied.  Point-spread function (PSF)
photometry was performed using {\sc iraf}'s {\sc daophot} task
\citet{stetson87}.  It was found that stars subtracted from an image using a
fixed PSF showed residuals that varied with position.  These residuals were
best reduced by employing {\sc daophot}'s ``penny2'' function and allowing it
to vary quadratically with position.  This is a two-component model, consisting
of an elliptical Gaussian core and Lorentzian wings.  Both parts of the model
are aligned along separate and arbitrary position angles.  {\sc daophot}
handles pixel defects, cosmic rays etc. by employing a formula which reduces
the weights of pixels that do not converge towards the model as the fit is
calculated.  This is discussed in more detail in \citet{allstar}.  The
post-processing (discussed below) is also able to detect and remove strongly
outlying points.  Where a star of interest is found to lie close to dead
columns/pixels conclusions have been drawn with caution.  We chose to have the
star positions re-fitted independently in each frame, having found that the
cross-correlation technique aligned the star centroids to around $\sim1$\,pixel
accuracy.  

For a significant number of images, we found that a position-dependent element
still remained in the magnitude residuals, particularly dominant along the long
($y$) axis of the CCD.  To counteract this problem, we have employed our own
post-processing software, described in \citet{street02}.  This included a
procedure which cross-correlates all lightcurves in the sample in order to
identify and remove remaining systematic trends.  

\begin{figure*}
\def\subfigtopskip{4pt}
\def\subfigbottomskip{8pt}
\def\subfigcapskip{4pt}
\centering
\begin{tabular}{cc}

\subfigure[CCD~1]{\label{fig:RMS1}
\includegraphics[angle=270.0,width=8.5cm]{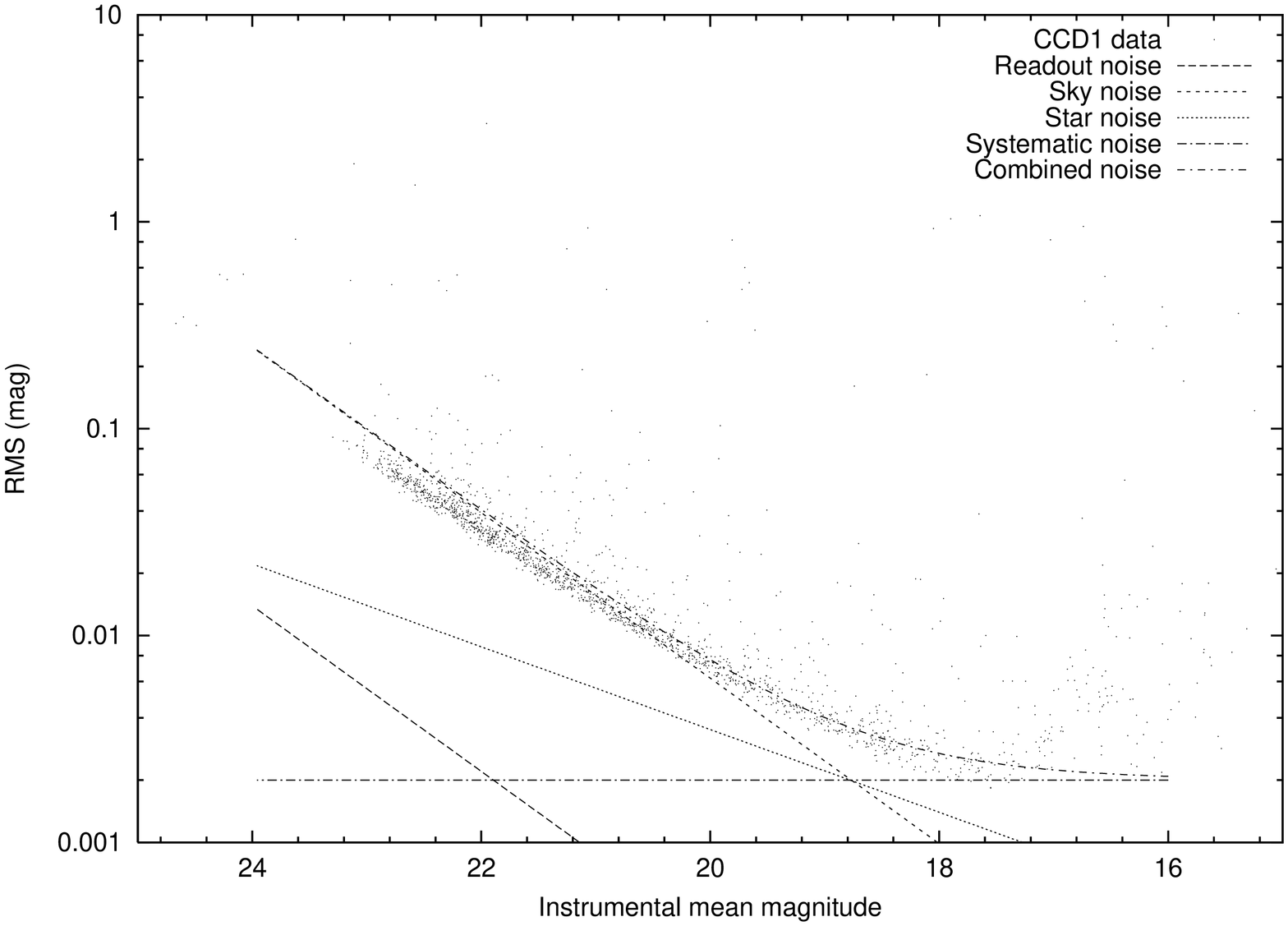}}
&
\subfigure[CCD~2]{\label{fig:RMS2}
\includegraphics[angle=270.0,width=8.5cm]{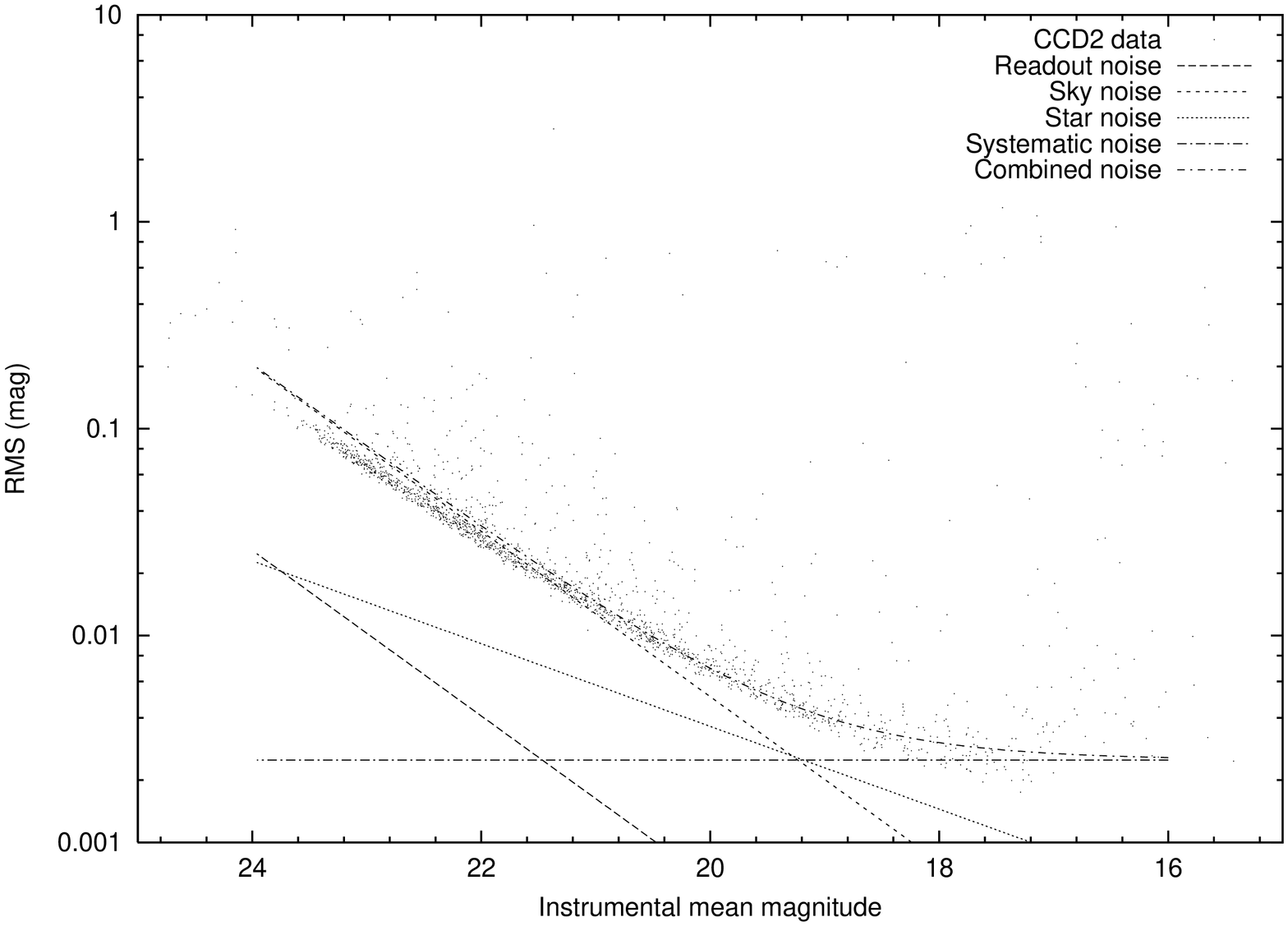}} \\

\subfigure[CCD~3]{\label{fig:RMS3}
\includegraphics[angle=270.0,width=8.5cm]{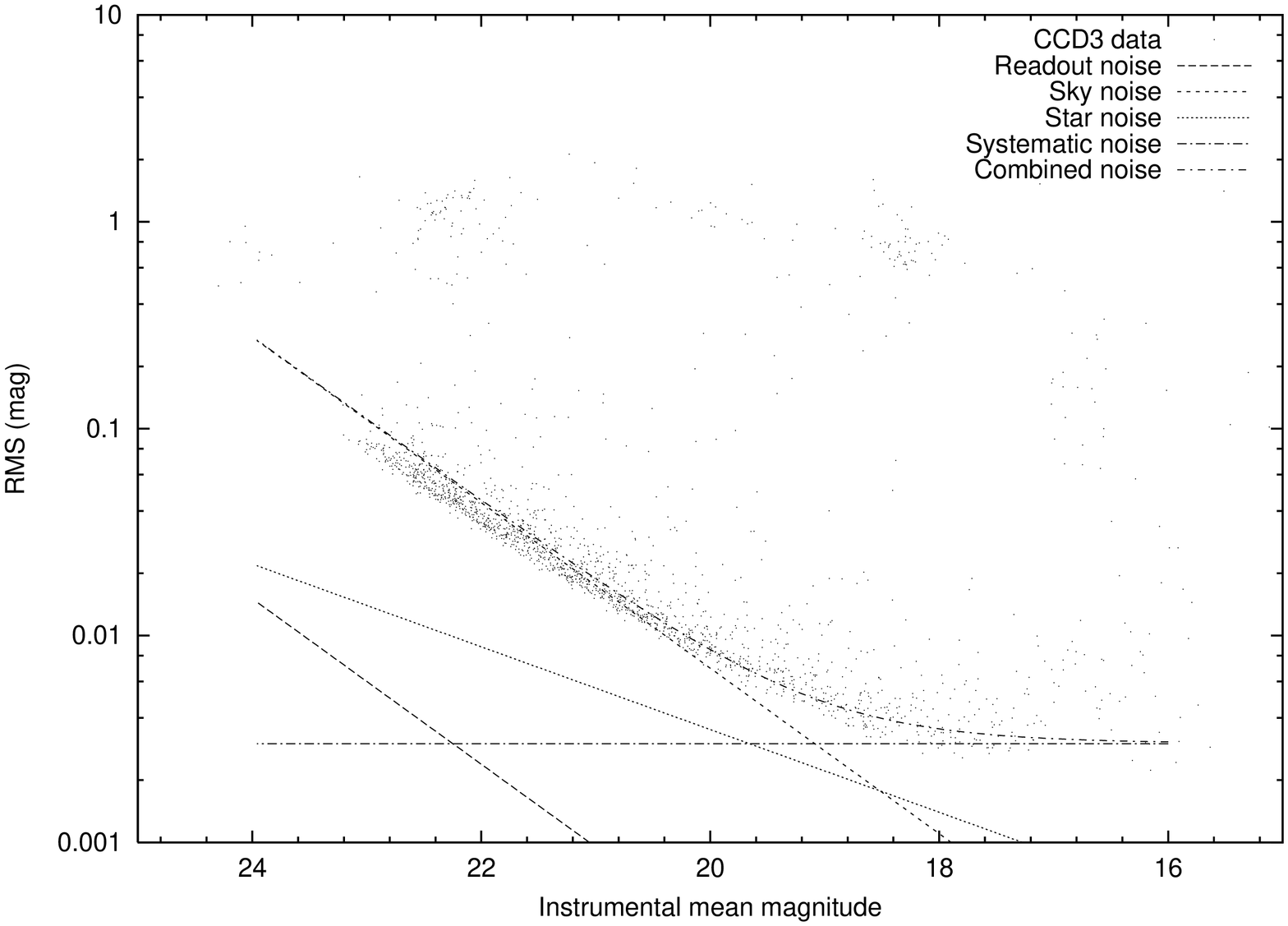}}
&
\subfigure[CCD~4]{\label{fig:RMS4}
\includegraphics[angle=270.0,width=8.5cm]{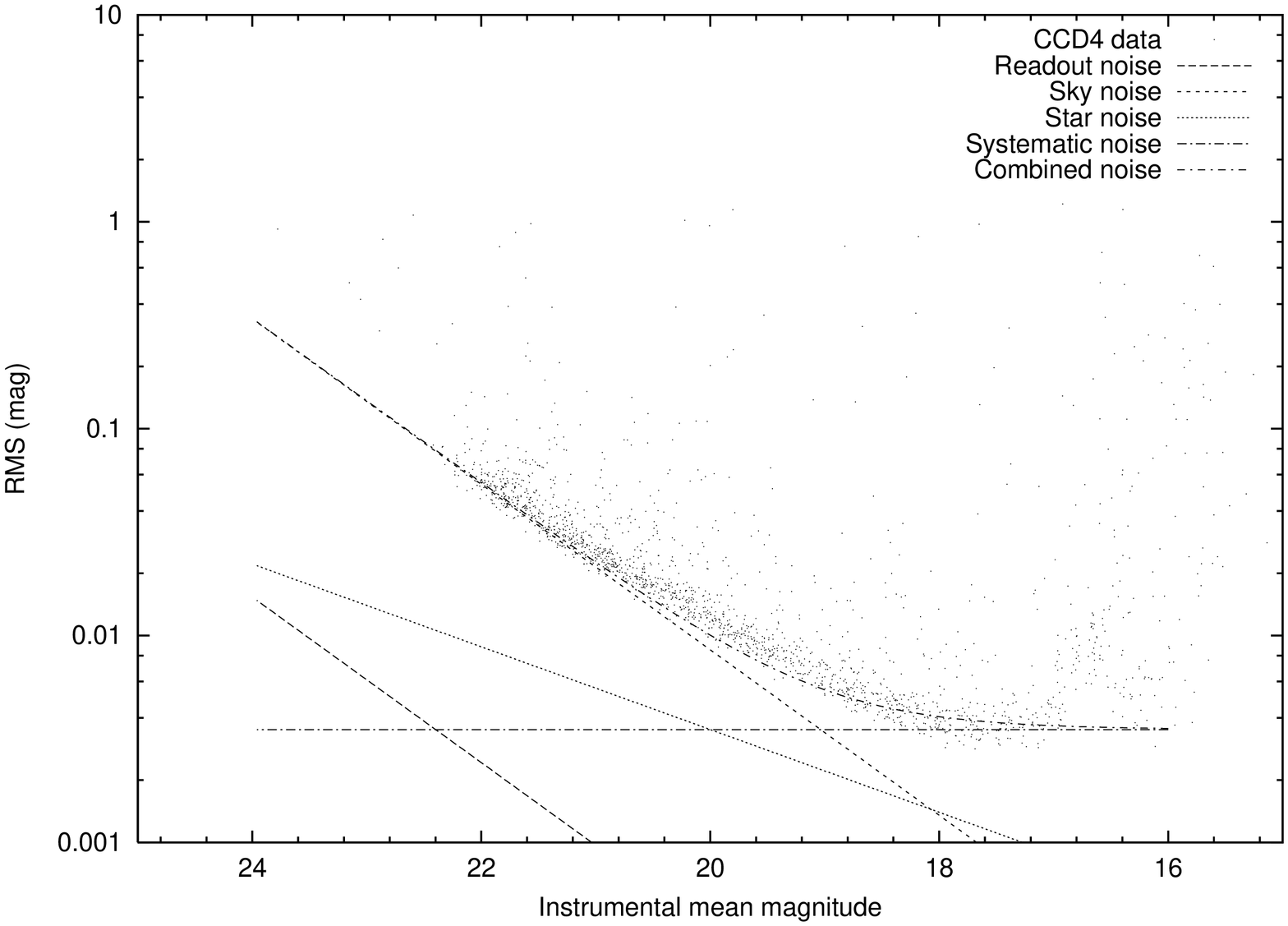}} \\

\end{tabular}
\caption{The variation of RMS scatter in star lightcurves with magnitude for
each of the four CCDs.  Superimposed curves show the effects of various sources
of noise.}
\protect\label{fig:RMSplot}
\end{figure*}

Following this processing, the precision achieved is illustrated by plotting
the RMS scatter in each star's lightcurve against its weighted mean magnitude
over the whole dataset.  Figure~\ref{fig:RMSplot} shows these plots for each
CCD, and for reference shows the effects of the main expected sources of
noise.  While some systematic effects remain in the data, our software improves
the precision particularly at brighter magnitudes, where we can achieve the
$\sim$0.004 mag precision required to detect planetary transits.  We notice
that the residual systematic variations are reduced to a level of
$\sim$0.0025--0.0035 mags.  After post-processing, these residuals do not
appear to show a positional distribution.  We also notice that the ``backbone''
of points falls slightly below the theoretical noise prediction at the faint
end.  This seems to be due to {\sc daophot} underestimating the magnitude
errors used to weight the calculations at fainter magnitudes.  We note the
presence of ``clumps'' of stars with high RMS in CCD3.  Investigation of these
points reveals that they are hot pixels located in the vignetted areas and dead
columns which this CCD suffers from.  

Astrometric positions for all the stars in our sample were obtained using the
method described in \citet{street02}.  The average RMS error in the resulting
RA and Dec are presented in Table~\ref{tab:astrom}, and correspond to an RMS
scatter of less than 1 pixel on the CCD.  

\begin{table} 
\centering 
\caption{Average RMS error on RA and Dec positions in arcsec for each CCD.}
\protect\label{tab:astrom} 
\vspace{5mm} 
\begin{tabular}{ccc} 
\hline 
CCD No.&   $\delta$RA	 &   $\delta$Dec   \\
       &   (arcsec)	 &   (arcsec)	   \\
\hline
1      &   0.118	 &   0.110	   \\
2      &   0.232	 &   0.363	   \\
3      &   0.289	 &   0.294	   \\
4      &   0.172	 &   0.180	   \\
\hline
\end{tabular} 
\end{table}

\section{Stellar Radii} 
\protect\label{sec:colours} 

\subsection{Colour Index Calibration}

\citet{kalirai01b} kindly provided us with $B-V$ colours indices for many of
the stars in the NGC~6819 field.  These data and the procedure used to
cross-identify stars are described in \citet{street02}.  Although most of our
data were taken in the Sloan r band, we obtained enough Sloan i data to
calibrate approximate $V-R$ colours for all the stars in our sample in the
following way.  We calculated the mean broadband flux using known passband
functions and the Bruzual, Persson, Gunn and Stryker atlas of stellar spectra. 
The mean broadband flux was then used to calculate theoretical magnitudes and
colours for a range of stellar spectral types relative to the flux from Vega in
that passband.  The INT instrumental Sloan $r-i$ colours were calibrated by
superimposing the {\sc xcal} plot of Sloan $r-i$ against $B-V$ over that of the
INT stars, and applying vertical and horizontal offsets.  The horizontal offset
provided the calibration factor for the INT data, in the sense that the true
Sloan $r-i$ colour of a star ($r-i$) is found from the instrumental one
($r-i)_{inst}$ by adding the offset, $\Delta_{r-i}$:

\begin{equation}
r-i = (r-i)_{inst} + \Delta_{r-i}.
\end{equation}

$\Delta_{r-i}$ was found to be 0\fm50, 0\fm50, 0\fm30 and 0\fm41 for the data
from CCDs 1, 2, 3 and 4 respectively.  In all cases, the vertical offset (the
difference between the theoretical and measured $B-V$) was found to be 0\fm2. 
This we attribute to extinction in the direction of the cluster and is not very
different from the value of $E(B-V) = 0.1$ measured by \citet{kalirai01}.

We used a similar method to convert the now-calibrated Sloan $r-i$ colours into
Johnson $V-R$.  {\sc xcal} was used to produce a dataset of $V-R$ and
corresponding Sloan $r-i$ values.  To derive a formula to convert Sloan $r-i$
colours into Johnson $V-R$, a function was fitted to these data using the
method of least squares.  As the shape of the curve changes at $r-i \sim 0.4$,
two functions were fitted; a straight line for $r-i$ values $\sim-0.2\sim0.4$
and an exponential function for the remaining curve:

\begin{eqnarray}
V-R = 
\begin{cases}
0.009 + 0.93(r-i)     	    & -0.2 \leq r-i \leq 0.356\\
1.30 - 1.72e^{-1.6(r-i)}    & 0.356 \leq r-i \leq 2.5
\end{cases}
\protect\label{eqn:VRcalib}
\end{eqnarray}

These relations were then used to calculate $V-R$ colour indices for all INT
stars with Sloan $r-i$ colours.  

\subsection{Colour-Magnitude Diagram}

\begin{figure*}
\def\subfigtopskip{4pt}
\def\subfigbottomskip{8pt}
\def\subfigcapskip{4pt}
\centering
\begin{tabular}{cc}

\subfigure[CCD~1]{\label{fig:HR1}
\includegraphics[angle=270.0,width=8.5cm]{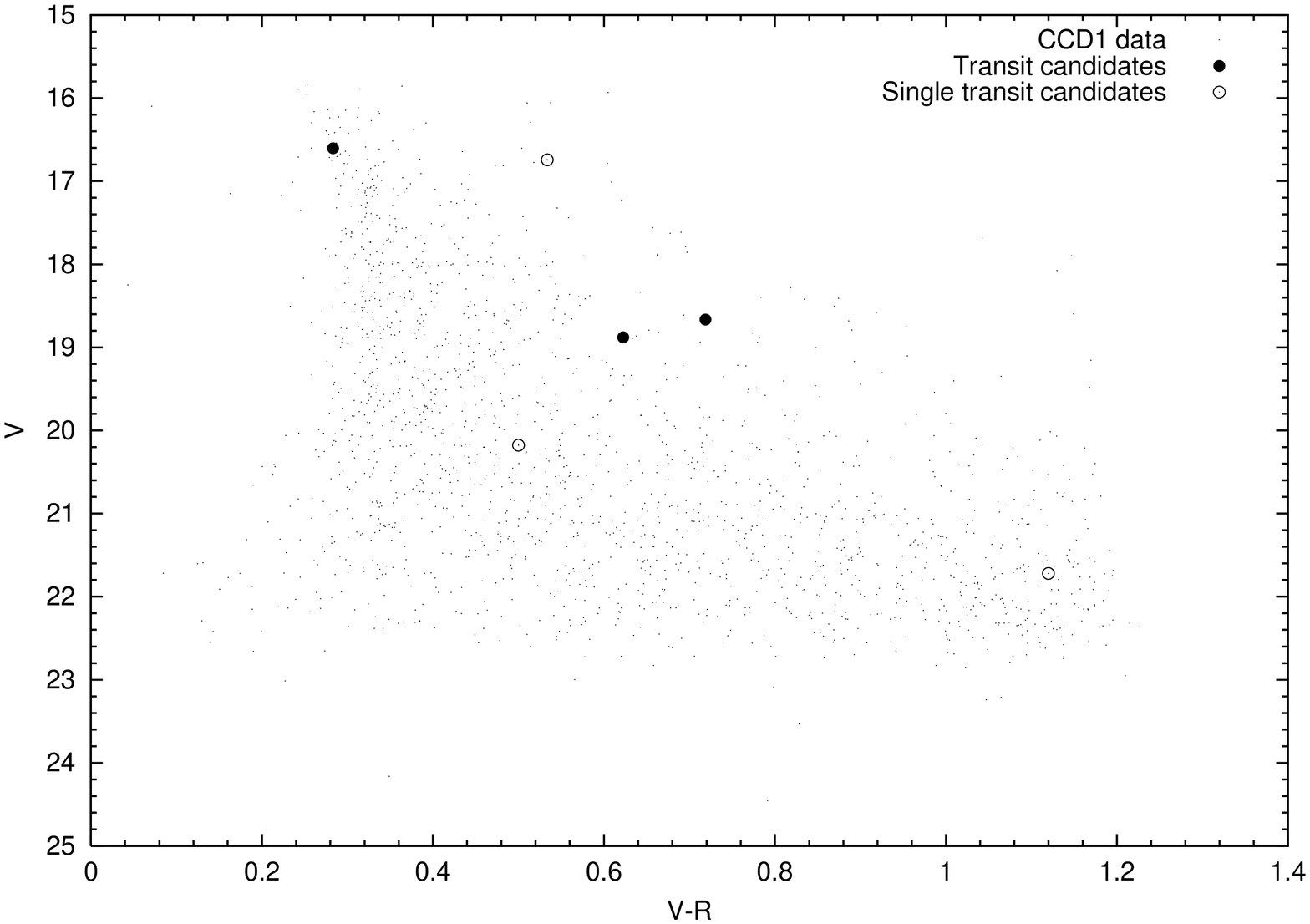}}
&
\subfigure[CCD~2]{\label{fig:HR2}
\includegraphics[angle=270.0,width=8.5cm]{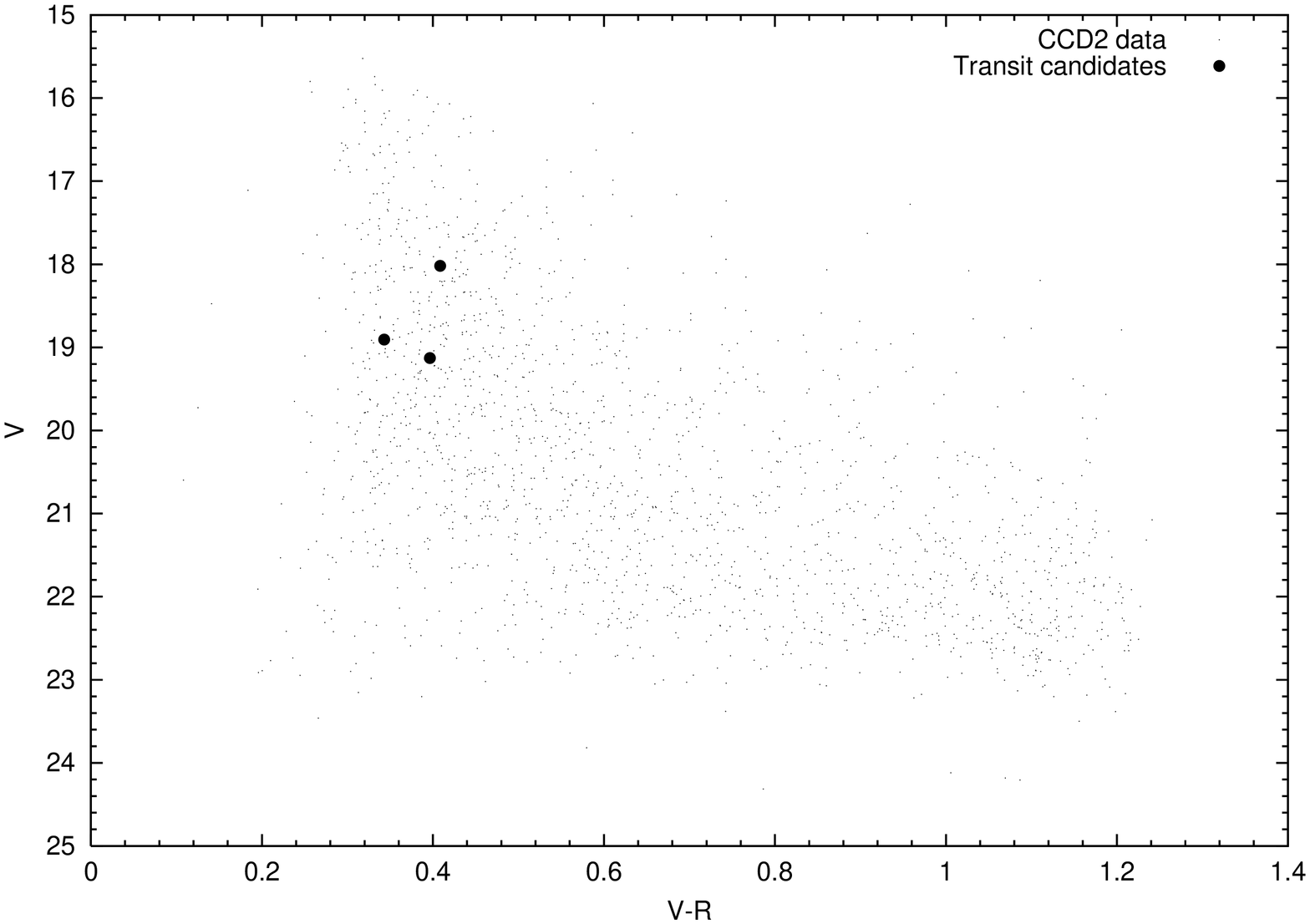}} \\

\subfigure[CCD~3]{\label{fig:HR3}
\includegraphics[angle=270.0,width=8.5cm]{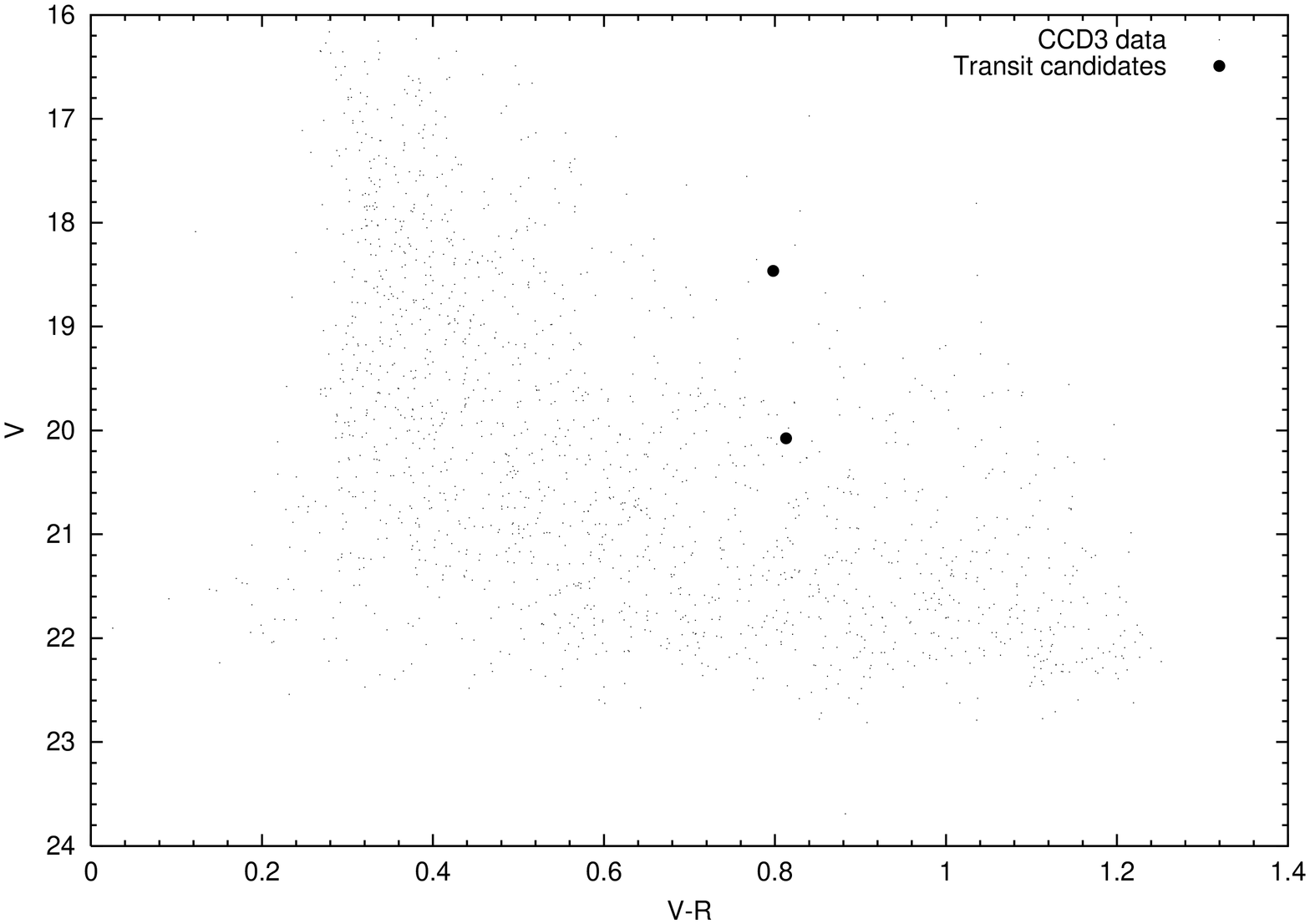}}
&
\subfigure[CCD~4]{\label{fig:HR4}
\includegraphics[angle=270.0,width=8.5cm]{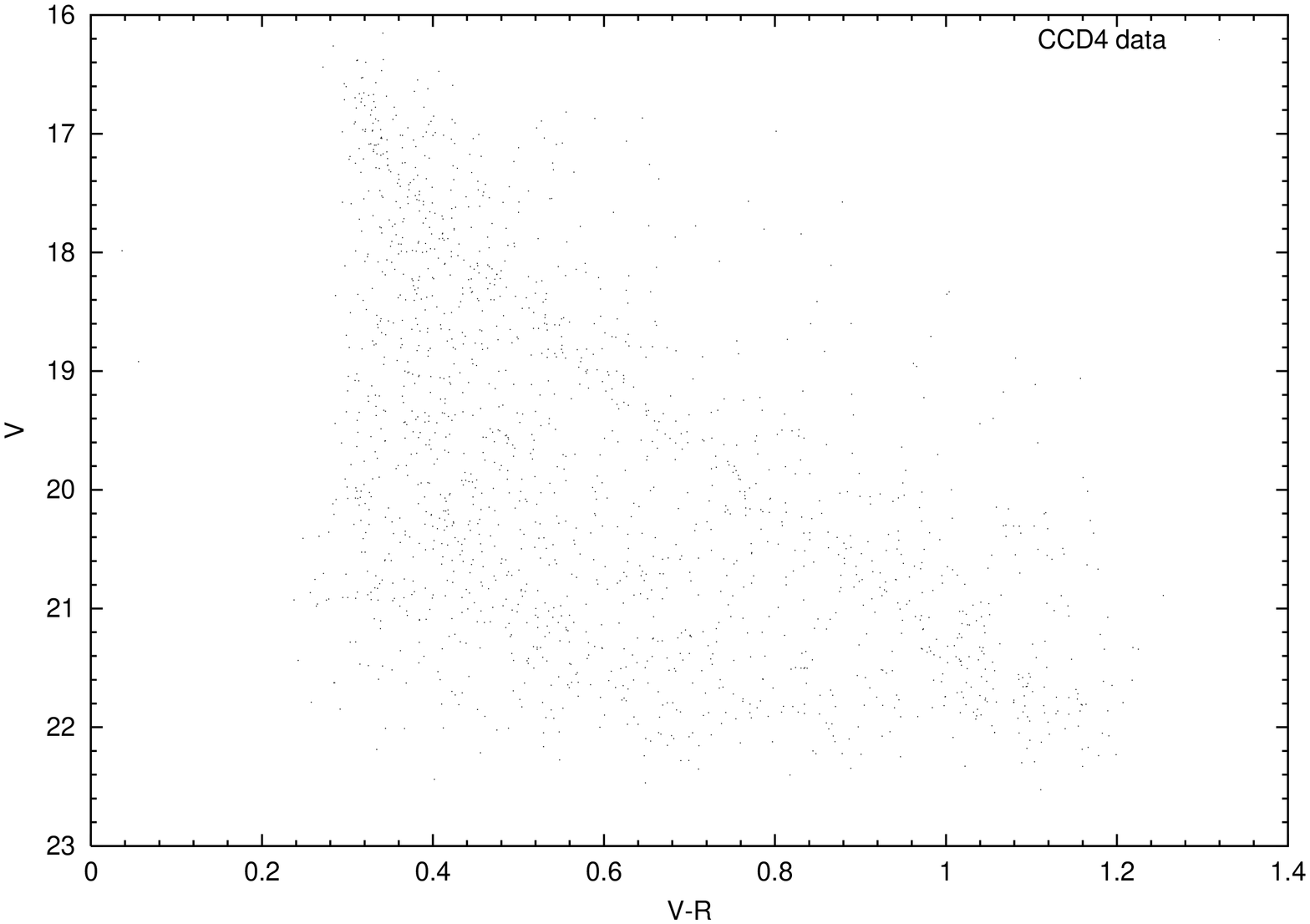}} \\

\end{tabular}
\caption{$V$.vs.$V-R$ colour-magnitude diagrams for the four WFC CCDs.  The
transit candidate stars are highlighted.}
\protect\label{fig:colourmag}
\end{figure*}

Figure~\ref{fig:colourmag} shows the $(V, V-R)$ colour-magnitude diagrams (CMD)
for the four CCDs.  The cluster main sequence is clearly visible in the data
from CCD4 and faintly in the other three plots.  This is expected from the
radius of the cluster \citet{kalirai01} which fits within the field of view of
one of the WFC CCDs.  Field stars greatly outnumber cluster members in these
data, and are located above and below the cluster main sequence in the CMD. 
The field stars above and below the cluster main sequence are predominantly
main sequence stars at closer and more remote distances than the cluster,
respectively.  In our following analysis, we assume that all stars in our
sample are main sequence.  This is reasonable since class IV subgiants are
sufficiently rare that their frequency in the sample is negligible.  Giant
stars are so bright ($-0\fm4 < M_{V} < 1\fm2$) that for one to be measurable
(unsaturated) in our data it would have to be at a distance of $\geq$9\,Kpc. 
We therefore adopt main sequence relationships and use the likely distance and
spectral type and hence radius of the stars in our sample.  

\begin{figure*}
\centering
\begin{tabular}{c}
\includegraphics[angle=270.0,width=8.5cm]{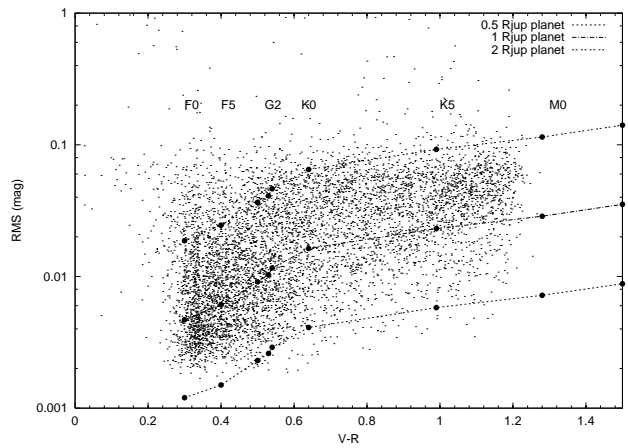}
\end{tabular}
\caption{RMS scatter against $V-R$ colour for all $\sim$38,000 stars in our
sample.  The overlaid curves give the predicted transit depth for cluster member
stars transited by planets of various radii.}
\protect\label{fig:RMScolour}
\end{figure*}

\subsection{Stellar Radii}

We use the $V-R$ index to estimate the radius of each star in this survey. 
This was done by interpolating between measured values of $V-R$ and $R_{*}$ for
the range of main sequence star types from \citet{gray92}.  These values were
supplemented at the low-mass end by data from \citet{reid97}, who list $V-R$
index, absolute magnitude and spectral type for 106 low-mass systems.  The
absolute magnitudes were then used to calculate the radii of these stars.  An
exponential curve was least-squares fitted to the Reid and Gizis data and used
to calculate values of stellar radii at fixed intervals of $V-R$ in order to
provide one smooth, continuous dataset.  Interpolation over this dataset was
then used to compute main sequence star radii from the $V-R$ colour index of
each star.  Figure~\ref{fig:RMScolour} plots RMS scatter versus $V-R$ colour. 
This is overlaid with curves illustrating the predicted transit depth of
planets with radii of 0.5, 1.0 and 2.0\,\Rj orbiting cluster member stars of
various masses.  If the RMS of a given star falls below one of these curves,
then we would expect to be able to detect a transit of a planet that size
around that star.  

We find that $\sim$30\% of the stars in our sample fall below the
1.0\,\Rj-transit curve, while $\sim$79\% fall below the 2.0\,\Rj line.  Thus
around $\sim$11,500 and $\sim$30,000 stars, respectively, are measured to
sufficient precision to allow the detection of transits.  Assuming $\sim$1\% of
all main sequence stars have hot Jupiter companions and $\sim$10\% of those
transit, then we can roughly expect to detect $\sim$11 Jupiter-radius objects
in our data.  

\section{Transit Detection Algorithm} 
\protect\label{sec:tf} 

Having obtained the required high-precision photometry, there are a number of
different approaches to the problem of detecting transit events.  We have
developed our own transit-finding software, using the method of matched-filter
analysis.  

After identifying and removing known large-amplitude variables from the data,
the software works in two stages.  The first stage, or ``standard search'',
generates a series of model lightcurves with a single transit.  These models
are generated for a range of transit durations $0.5 < \delta t < 5.0$\,hours,
in intervals of 0.25\,hours and with the time of mid-transit ranging from the
start to the end of the observing campaign in steps of $\delta t/4$.  A
constant magnitude is least-squares-fitted to each lightcurve and the
corresponding $\chi^{2}_{c}$ is calculated.  Each model is then 
$\chi^{2}$-fitted to each lightcurve, the transit depth and out-of-transit
magnitude being optimised by minimising the quality-of-fit statistic
$\chi^{2}_{m}$.  The details of the best-fitting model with the lowest value of
$\chi^{2}_{m}$ are stored.  

A transit-finder index, $\Delta \chi^{2}_{tf}$, is then calculated as:

\begin{equation}
\protect\label{eqn:tfindex}
\Delta \chi^{2}_{tf} = \chi^{2}_{c} - \chi^{2}_{m}.  
\end{equation}

\begin{figure*}
\centering
\begin{tabular}{c}
\includegraphics[angle=270.0,width=8.5cm]{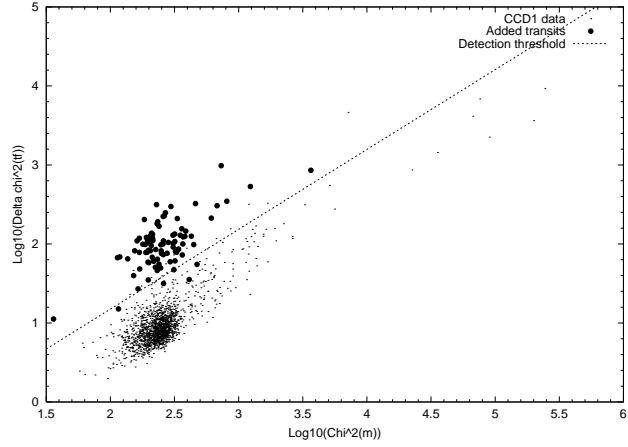}
\end{tabular}
\caption{Plot of $\Delta \chi^{2}_{tf}$ against $\chi^{2}_{m}$ for the CCD1
dataset with fake HD~209458-like transits added; these points are highlighted.  
Transits fall to the upper left of this diagram, above the main backbone of 
points -- the threshold is set at $+4\sigma$.}
\protect\label{fig:addedtrans}
\end{figure*}

Plotting this index against $\chi^{2}_{m}$ allows us to separate transit events
from constant stars and other types of variables.  A genuine transit event
would be expected to show a significant improvement in $\chi^{2}$ when
comparing the fit of a constant line and a suitable transit model; hence it
would have a relatively low value of $\chi^{2}_{m}$ and a high value of $\Delta
\chi^{2}_{tf}$.  To isolate these candidates, a straight line is least-squares
fitted to the ``backbone'' of points.  This fit is iterated, rejecting all
points $\pm3\sigma$ above the line, until the parameters change by less than
0.0001.  A cut-off line is established by raising this line by $+N\sigma$ where
$N$ is set by the user; all stars falling to the top-left of this cut-off are
regarded as candidates.  To illustrate how this isolates transit candidates, we
tested the algorithm by injecting fake transits with known parameters into the
data stream.  Transits were added to 1\% of stars in the CCD1 data, with a
period of 3.4\,days, a duration of 2.5\,hours, and an amplitude of 0\fm02.  The
first transit occurred at HJD 2451355.5 and at every multiple of the period. 
The transit-finder algorithm was then applied to this modified dataset and the
resulting plot of $\Delta \chi^{2}_{tf}$ against $\chi^{2}_{m}$ is shown in
Figure~\ref{fig:addedtrans}.  The transits of an HD~209458b-like planet are
clearly separated from the rest of the data.  This figure was used to set the
detection threshold.  It was found that a $+4\sigma$ threshold retains all but
5 of the 90 fake transits while excluding $\sim$99.2\% of the constant stars.  

The second stage of the search is a ``period search'' applied to candidates
highlighted by the standard search.  For each candidate, multiple-transit models
are generated across a range of periods (2--5 days) and fitted to the lightcurve
as described above.  Once again, the minimum value of $\chi^{2}_{m}$ sets the
best-fitting model, and candidates are selected by the method described above. 
This two stage approach ensures that all the relevant transit-parameters are
determined for all candidates, but restricts the number of least-squares fits
required by applying the period search to transit-candidates only.  This allows
a statistically-optimal matched-filter technique to be applied without
prohibitive computational time requirements.

\begin{figure*}
\def\subfigtopskip{4pt}
\def\subfigbottomskip{8pt}
\def\subfigcapskip{4pt}
\centering
\begin{tabular}{cc}

\subfigure[CCD~1]{\label{fig:chi1}
\includegraphics[angle=270.0,width=8.5cm]{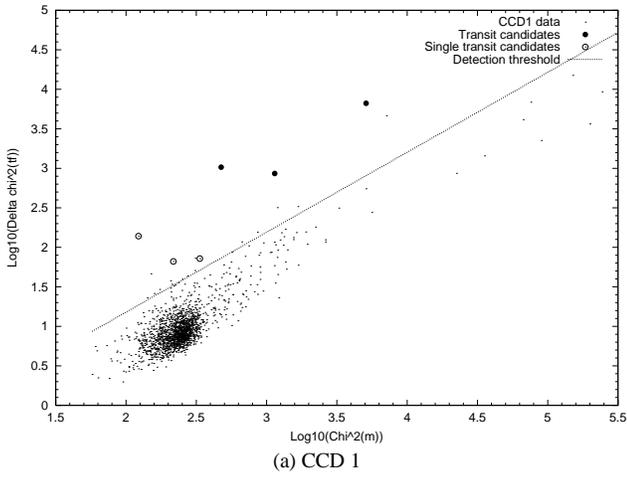}}
&
\subfigure[CCD~2]{\label{fig:chi2}
\includegraphics[angle=270.0,width=8.5cm]{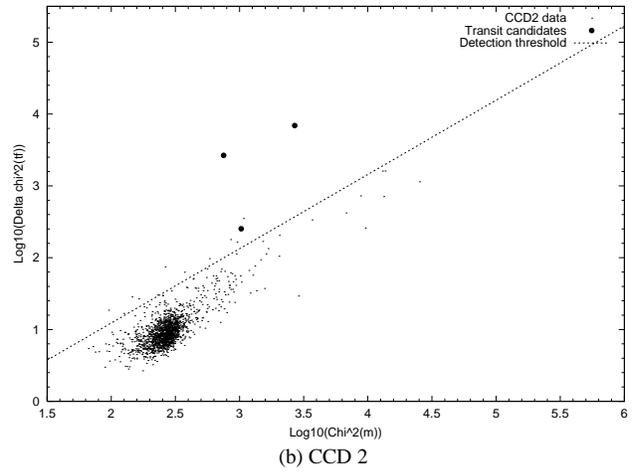}} \\

\subfigure[CCD~3]{\label{fig:chi3}
\includegraphics[angle=270.0,width=8.5cm]{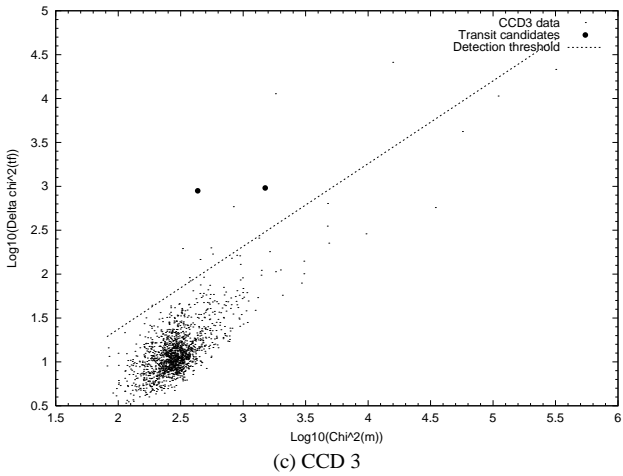}}
&
\subfigure[CCD~4]{\label{fig:chi4}
\includegraphics[angle=270.0,width=8.5cm]{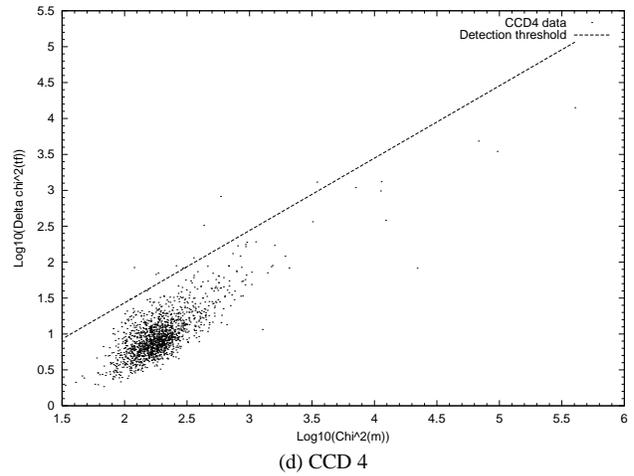}} \\

\end{tabular}
\caption{Plots of $\Delta \chi^{2}_{tf}$ against $\chi^{2}_{m}$ for all four
CCDs.  All lightcurves above the superimposed cut-off thresholds are
visually examined for transits. }
\protect\label{fig:transitid}
\end{figure*}

We note that an element of human judgement enters into our transit detection
procedure.  The candidates presented by the algorithm are sorted by manual
examination.  In the process we have rejected a large number of ``possible''
transits; lightcurves which show dips at the beginning or end of a night which
do not repeat, for example, or those that show dips sampled with very few
datapoints.  Of course this means we could potentially miss transit
ingresses/egresses, but a real candidate must show at least two well-sampled
transit events.  

\section{Results}
\protect\label{sec:results}

In total, over 38,000 star lightcurves have been analysed in this way.  The
transit search algorithm highlighted 276 stars worthy of further investigation,
and these were examined manually.  The majority ($\sim$51.9\%) were found to
show only a few, fainter-than-average, scattered points.  The cause of the
scattering was found to be one of three situations: (a) the presence of nearby
or blended companion star(s), (b) the star is bright and saturated in a
significant number of images or (c) the star falls close to a dead column or
vignetted area on the CCD.  The algorithm also detected what we judge to be
stellar variability in $\sim$20.3\% of cases; most of these stars showed
eclipses due to stellar companions while some displayed low-amplitude ``dips''
in brightness due to stellar activity, although on longer timescales than
transits.  No obvious explanation for spurious detection could be found for 67
of these stars; in these cases examination of the lightcurve revealed
unconvincing ``transits'' consisting of well-scattered points, often on nights
of poor conditions.  

Of the remaining stars ($\sim$2.9\%), 8 appear to show short-duration
transit-like eclipses.  The sample includes a number of active stars which show
brief eclipses.  This is not unexpected since transit amplitude scales
inversely with star radius squared, while stellar activity is more common among
small, young stars.  All short-duration eclipses were considered regardless of
amplitude, since hot Jupiter transits could reach depths of up to several
tenths of a magnitude, given a late-M type primary.  

A full lightcurve solution to a fitted model is not possible because of the
sparseness of the data.  However, the $V-R$ colours for these stars can be used
to estimate the radius of the primary star ($R_{*}$), assuming for the moment
that the star is main sequence and that negligible light is contributed from
the companion body.  The amplitude, $\delta m$, of the transit is proportional
to the ratio of the star's radius to that of the companion ($R_{c}$):

\begin{equation}
\delta m \approx \left ( \frac{R_{c}}{R_{*}} \right )^{2}, 
\protect\label{eqn:compradius}
\end{equation}

We estimate $R_{c}$ from Equation~\ref{eqn:compradius}, noting that this gives
a lower limit because a larger companion can cover the same fraction of primary
star if the eclipse is partial rather than total.  The radius of the companion
gives a general indication of its nature.  However, while the radius of a main
sequence M~star can be $\sim$0.1--0.5\,\Rsol, the radii of gas giant planets
(\Rj$\approx$0.1\,\Rsol) are thought to be similar to those of brown dwarfs
($\leq$0.2\,\Rsol), owing to their degenerate nature.  The transits of
HD~209458b have also shown that hot Jupiter radii can be larger than expected
due to their early proximity to their primary star slowing the rate of
contraction \citet{burrows00}.   For this reason, follow-up radial velocity
measurements yielding the minimum mass will be required to distinguish between
planetary and stellar companions.  

\begin{figure*}
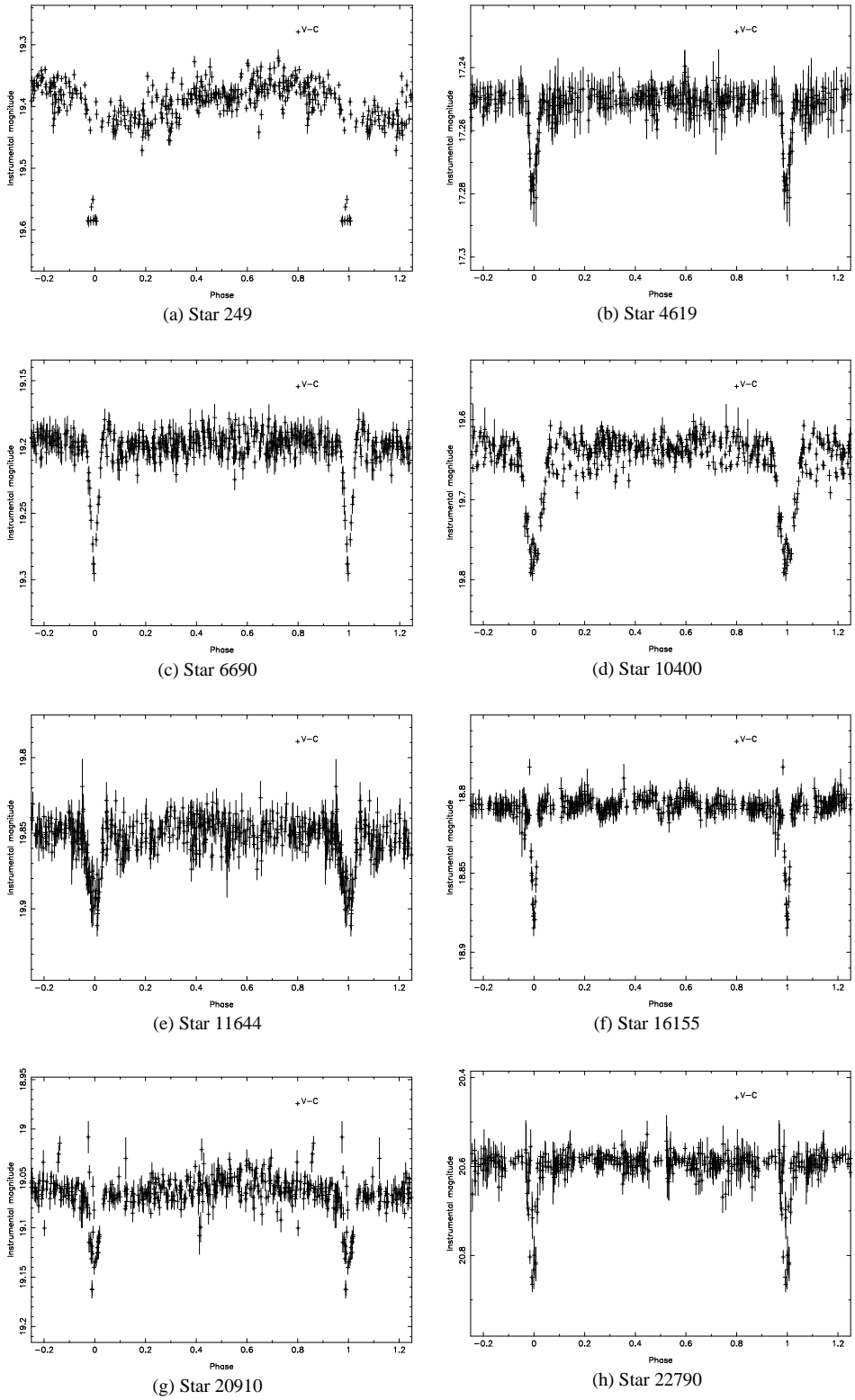

\def\subfigtopskip{4pt}
\def\subfigbottomskip{8pt}
\def\subfigcapskip{4pt}
\centering
\begin{tabular}{cc}

\subfigure[Star 249]{\label{fig:ph249}
\includegraphics[angle=270.0,width=6cm]{ph249.ps}}
&
\subfigure[Star 4619]{\label{fig:ph4619}
\includegraphics[angle=270.0,width=6cm]{ph4619.ps}} \\

\subfigure[Star 6690]{\label{fig:ph6690}
\includegraphics[angle=270.0,width=6cm]{ph6690.ps}} 
&
\subfigure[Star 10400]{\label{fig:ph10400}
\includegraphics[angle=270.0,width=6cm]{ph10400.ps}} \\

\subfigure[Star 11644]{\label{fig:ph11644}
\includegraphics[angle=270.0,width=6cm]{ph11644.ps}}
&
\subfigure[Star 16155]{\label{fig:ph16155}
\includegraphics[angle=270.0,width=6cm]{ph16155.ps}} \\

\subfigure[Star 20910]{\label{fig:ph20910}
\includegraphics[angle=270.0,width=6cm]{ph20910.ps}} 
&
\subfigure[Star 22790]{\label{fig:ph22790}
\includegraphics[angle=270.0,width=6cm]{ph22790.ps}} \\

\end{tabular}
\caption{The phase-folded lightcurves of stars showing transit-like eclipse events.}
\protect\label{fig:phasecurves1}
\end{figure*}

Table~\ref{tab:yield} presents the details of the 8 candidates.  Note that one
candidate is marked as being blended with nearby stars.  The conclusions drawn
from these stars come with the caveat that the results need to be confirmed. 
All the candidates are found to have a minimum companion radius below
0.5\,\Rsol, while 3 have $R_{c} \leq 0.25\,\Rsol$.  The phased lightcurves of
the stars with M-dwarf or smaller companions are presented in
Figure~\ref{fig:phasecurves1}.  The candidates are discussed individually
below.  Some of the companion objects could be brown dwarfs although most are
found to be low mass stars.  In these cases, we note that stellar companions
will contribute by reddening the measured colour - this would mean that the
primary and companion stars are of larger radii than calculated here, as would
inclinations of less than 90$^{\circ}$.   Stellar binaries would also exhibit
secondary eclipses not seen in transit lightcurves.   We have not included
objects which clearly show eclipses of different depths, as these will be
stellar binaries.  Where objects showing similar eclipse-depths turn out to be
stars, then the period will be twice that given in Table~\ref{tab:yield}.  

\subsection{Star 249 -- P=2.233\,days, $\delta m$=0\fm19} 

With all binary objects with periods as short as these it is possible for the
rotation of the primary to have become synchronised with the orbital period of
the companion.  If this is the case, then stellar activity on the primary is
expected, resulting in the variable out-of-eclipse lightcurves.  This seems to
be the case for star 249: the implied companion radius (0.37\,\Rsol) is that of
an M-dwarf or larger.  While the eclipses are not well sampled there is some
suggestion that they may be rounded-bottomed.  

\subsection{Star 4619 -- P=3.682\,days, $\delta m$=0\fm03}

This candidate shows the classic transit lightcurve: sharp ingress/egress to low
amplitude eclipse with no out of transit variations.  The eclipse profiles are
not well sampled but could be rounded-bottomed, and the period is typical of the
known hot Jupiters.  The $V-R$ colour (0\fm283) indicates a primary radius of
$\sim$1.5\,\Rsol which together with the low amplitude implies a companion
radius of $\sim$0.26\,\Rsol.  The companion could be a brown dwarf.  

\subsection{Star 6690 -- P=1.682\,days, $\delta m$=0\fm09}

This candidate also shows the expect transit lightcurve except that the eclipses
appear to have a sharp, pointed profile, suggesting that these are grazing
incidence eclipses.  This would mean that the companion radius is larger than
0.23\,\Rsol.  However, the colour (0\fm719) and amplitude (0\fm09) imply a
relatively small primary and secondary radii.  In this case the companion is
likely to be a low mass star.  

\subsection{Star 10400 -- P=1.46\,days, $\delta m$=0\fm14}

Measurements of this star are complicated by the presence of close, blended
companions.  The colour (0\fm343) and amplitude (0\fm14) indicate that both
primary and secondary radii are stellar.  

\subsection{Star 11644 -- P=2.302\,days, $\delta m$=0\fm04}

The lightcurve of this star also shows some modulation between the eclipses,
which seem to be flat-bottomed; further photometric data are needed to confirm
this.  The colour (0\fm397) and amplitude (0\fm04) suggest that the primary has
a radius of $\sim$1.32\,\Rsol while the minimum secondary radius is found to be
$\sim$0.264\,\Rsol.  The companion object could therefore be a brown dwarf.  

\subsection{Star 16155 -- P=3.486\,days, $\delta m$=0\fm07}

This lightcurve is similar to that of star 6690, with pointed eclipse
profiles.  However, in this case even the companion's minimum radius
(0.35\,\Rsol) implies a small star and a grazing incidence eclipse suggests a
larger companion.  

\subsection{Star 20910 -- P=1.3112\,days, $\delta m$=0\fm1}

This lightcurve displays eclipses apparently rounded and weak rotational
modulation of the out-of-eclipse lightcurve.  The 1.3\,day period and early-K
spectral type suggest that magnetic starspot activity driven by the
tidally-synchronised rotation of the primary is responsible for the
modulation.  The lack of a clear secondary eclipse suggests a very low
effective temperature for the companion, which is probably a late-M dwarf or a
brown dwarf.  

\subsection{Star 22790 -- P=3.621\,days, $\delta m$=0\fm25}

The eclipse profile again suggests a grazing incidence orbit while the minimum
radius (0.37\,\Rsol) implies that the companion is a low mass star.  

\begin{table*}
\centering
\caption{The parameters of the stars which show transit-like eclipses.}
\protect\label{tab:yield}
\vspace{5mm}
\begin{tabular}{cccccccccccc}
\hline
Star	  & $V$    & $V-R$  & $\delta m$  & $\delta t$& $R_{*}$ & $R_{c}$ & Period    & Epoch	      & $N_{tr}$ & RA	   & Dec	 \\
	  & (mag)  & (mag)  & (mag)	  & (hours)   & (\Rsol) & (\Rsol) & (days)    & (HJD-2400000)&        	& (J2000.0)   & (J2000.0)  \\
\hline
249   	  & 18.879 & 0.623  & 0.19	  & 2.4       &  0.84   & 0.37	  & 2.233(31) & 51352.535(1) & 2      	& 19 42 15.05 & +40 04 42.1\\
4619	  & 16.603 & 0.283  & 0.03	  & 4.8       &  1.51   & 0.26	  & 3.682(1)  & 51387.609(4) & 2      	& 19 41 21.32 & +40 02 14.3\\
6690  	  & 18.667 & 0.719  & 0.09	  & 3.1       &  0.77   & 0.23	  & 1.682(1)  & 51356.578(7) & 3      	& 19 40 56.71 & +40 05 05.0\\
10400$^a$ & 18.906 & 0.343  & 0.14        & 4.3       &  1.16   & 0.43    & 1.46(6)   & 51357.506(1) & 5      	& 19 40 05.30 & +40 14 17.7\\
11644 	  & 19.130 & 0.397  & 0.04    	  & 3.6       &  1.32 	& 0.26	  & 2.302(2)  & 51382.534(4) & 3.5      & 19 40 13.93 & +40 11 21.9\\
16155 	  & 18.018 & 0.408  & 0.07	  & 2.6       &  1.32   & 0.35	  & 3.486(5)  & 51359.526(2) & 3      	& 19 40 12.44 & +40 00 45.1\\
20910 	  & 18.464 & 0.798  & 0.10	  & 1.9       &  0.73   & 0.23	  & 1.3112(6) & 51356.4832(9)& 5      	& 19 41 57.10 & +40 18 25.3\\
22790 	  & 20.075 & 0.813  & 0.25	  & 4.6       &  0.73   & 0.37	  & 3.621(2)  & 51383.490(15)& 2      	& 19 41 33.86 & +40 26 35.0\\
\hline
\multicolumn{10}{l}{\parbox{14cm}{\small $^a$ Blended}}
\end{tabular}
\end{table*}

\subsection{Single-transit Candidates}

\begin{figure*}
\def\subfigtopskip{4pt}
\def\subfigbottomskip{8pt}
\def\subfigcapskip{4pt}
\centering
\begin{tabular}{cc}

\subfigure[Lightcurve of star 829]{\label{fig:lc829}
\includegraphics[angle=270.0,width=8.5cm]{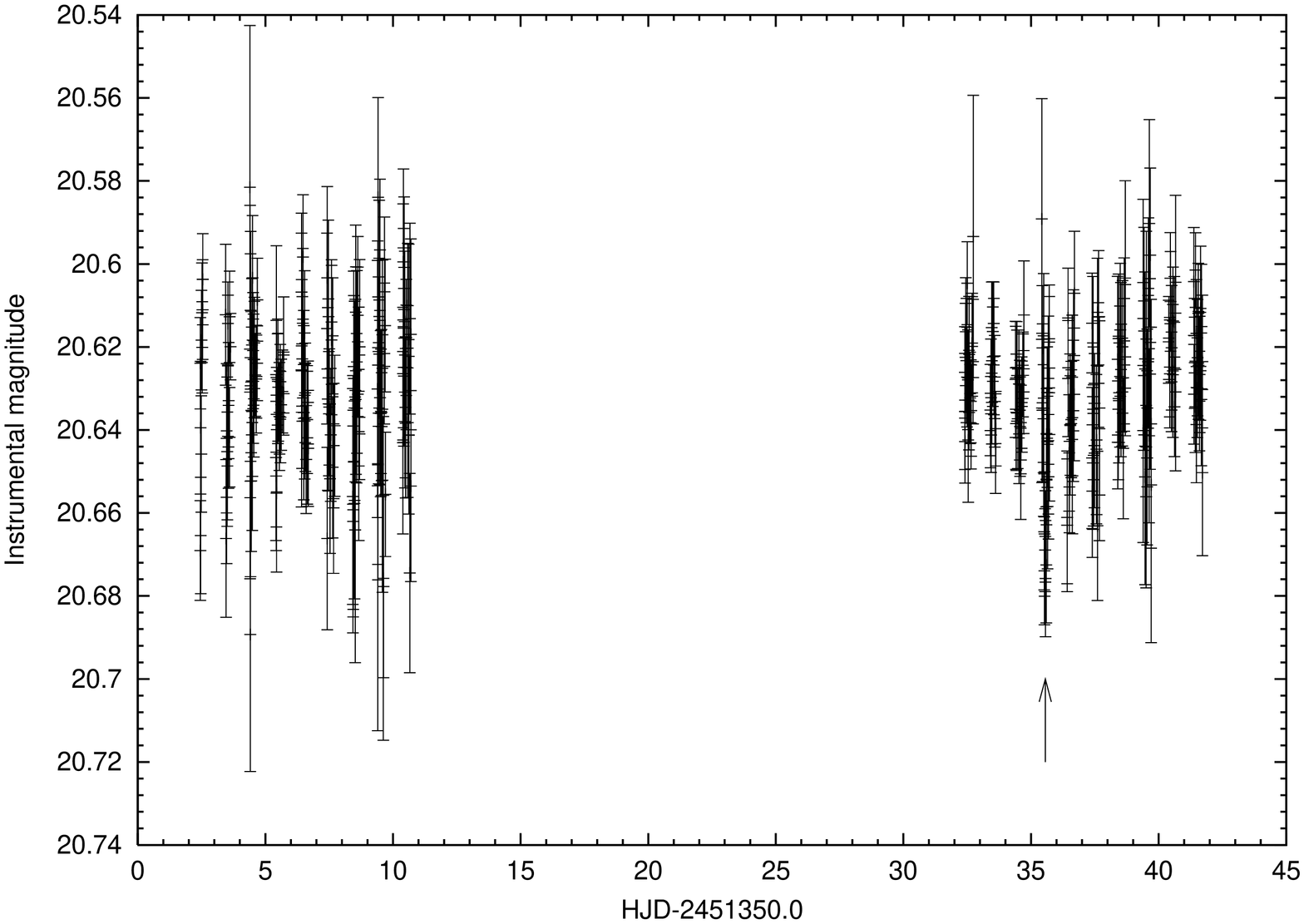}}
&
\subfigure[Star 829]{\label{fig:trans829}
\includegraphics[angle=270.0,width=8.5cm]{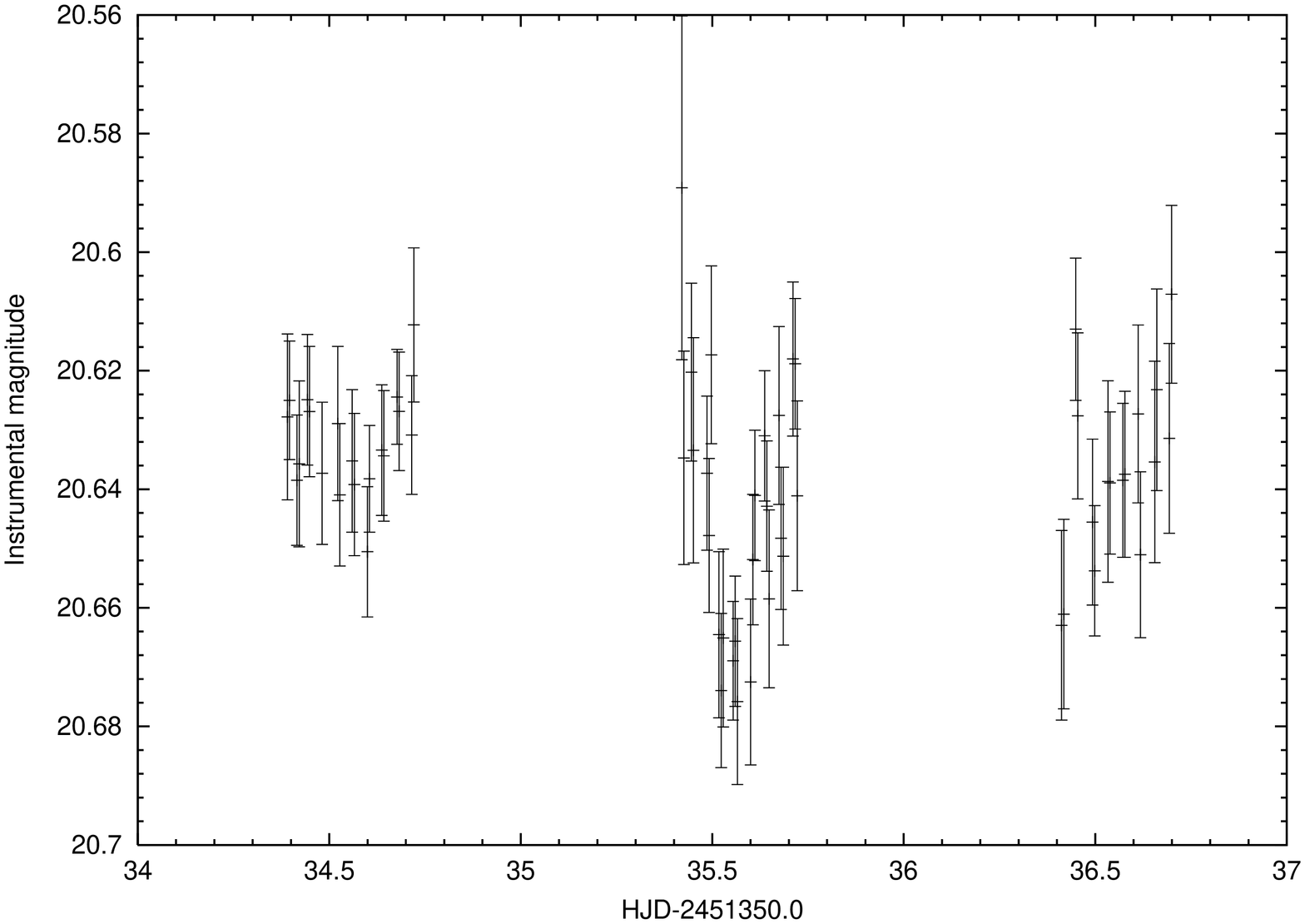}} \\

\subfigure[Lightcurve of star 8153]{\label{fig:lc8153}
\includegraphics[angle=270.0,width=8.5cm]{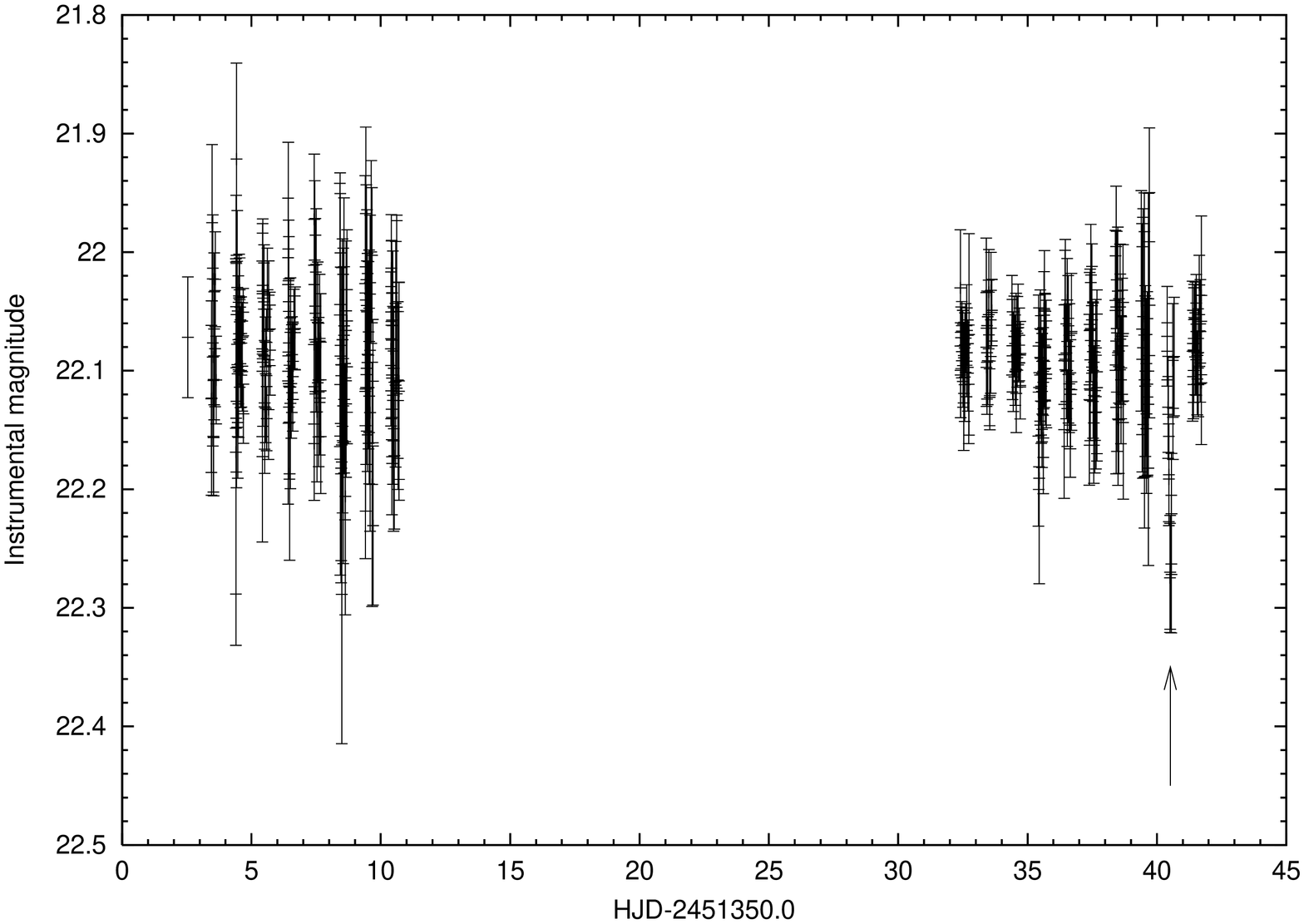}} 
&
\subfigure[Star 8153]{\label{fig:trans8153}
\includegraphics[angle=270.0,width=8.5cm]{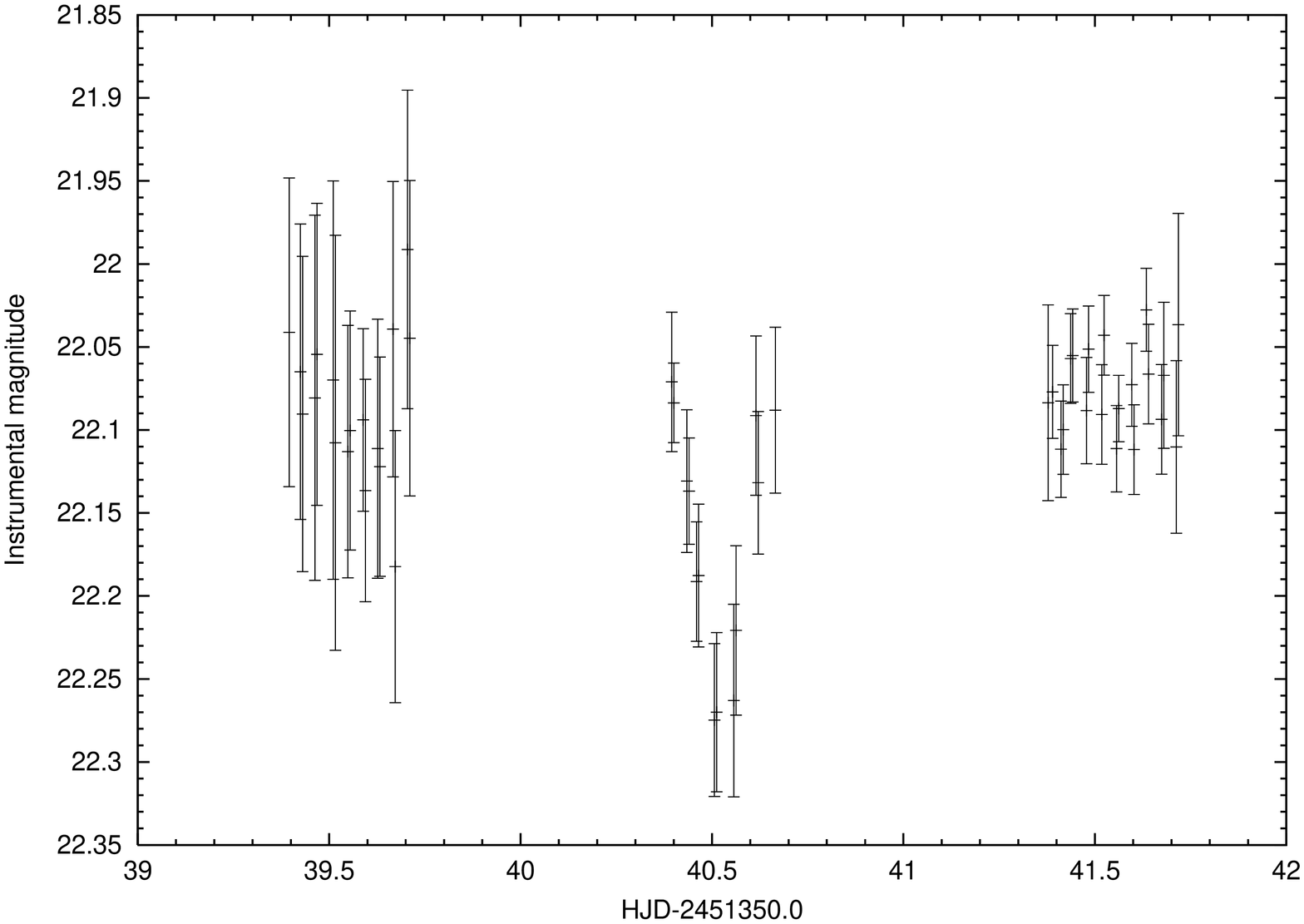}} \\

\subfigure[Lightcurve of star 9329]{\label{fig:lc9329}
\includegraphics[angle=270.0,width=8.5cm]{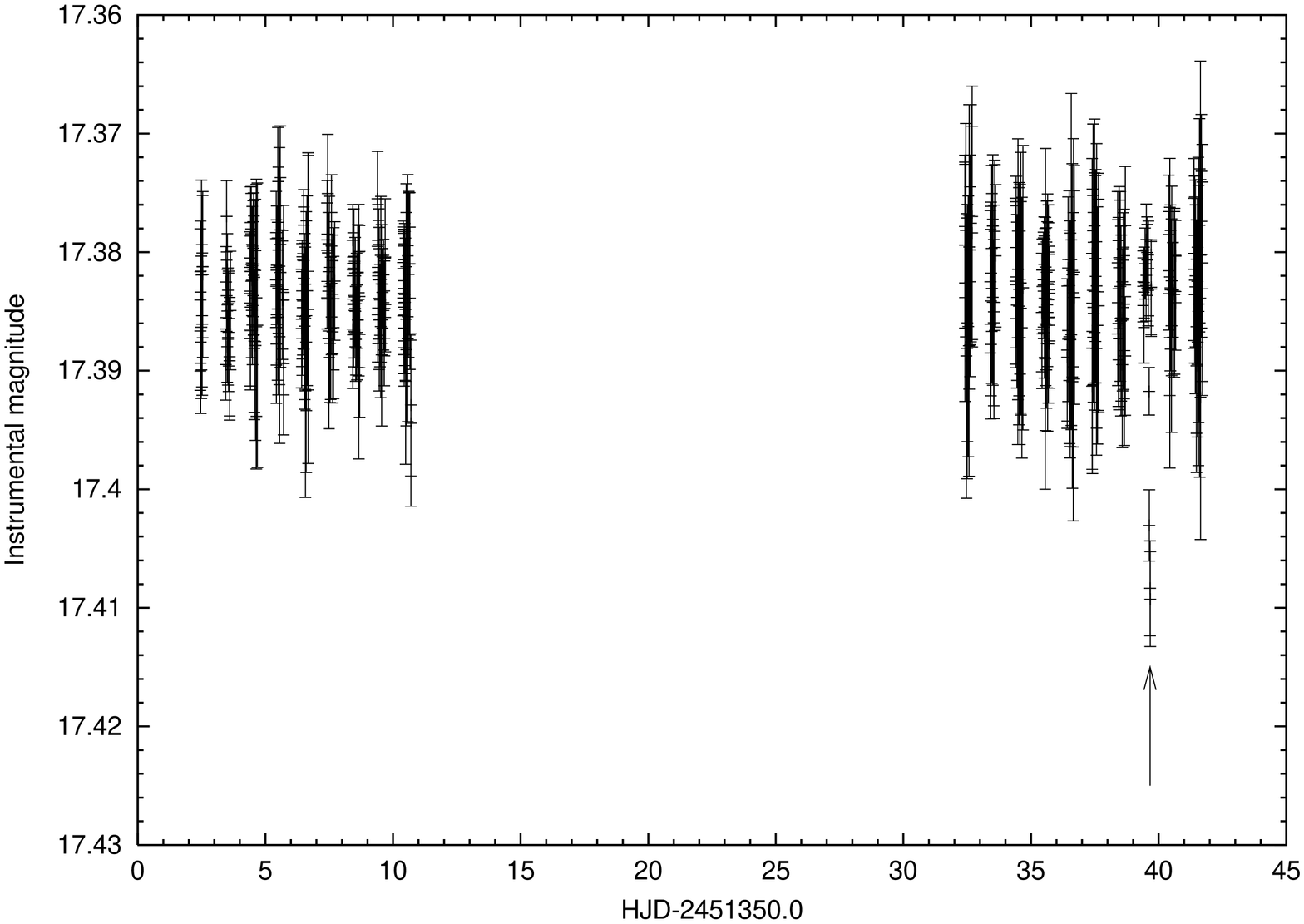}}
&
\subfigure[Star 9329]{\label{fig:trans9329}
\includegraphics[angle=270.0,width=8.5cm]{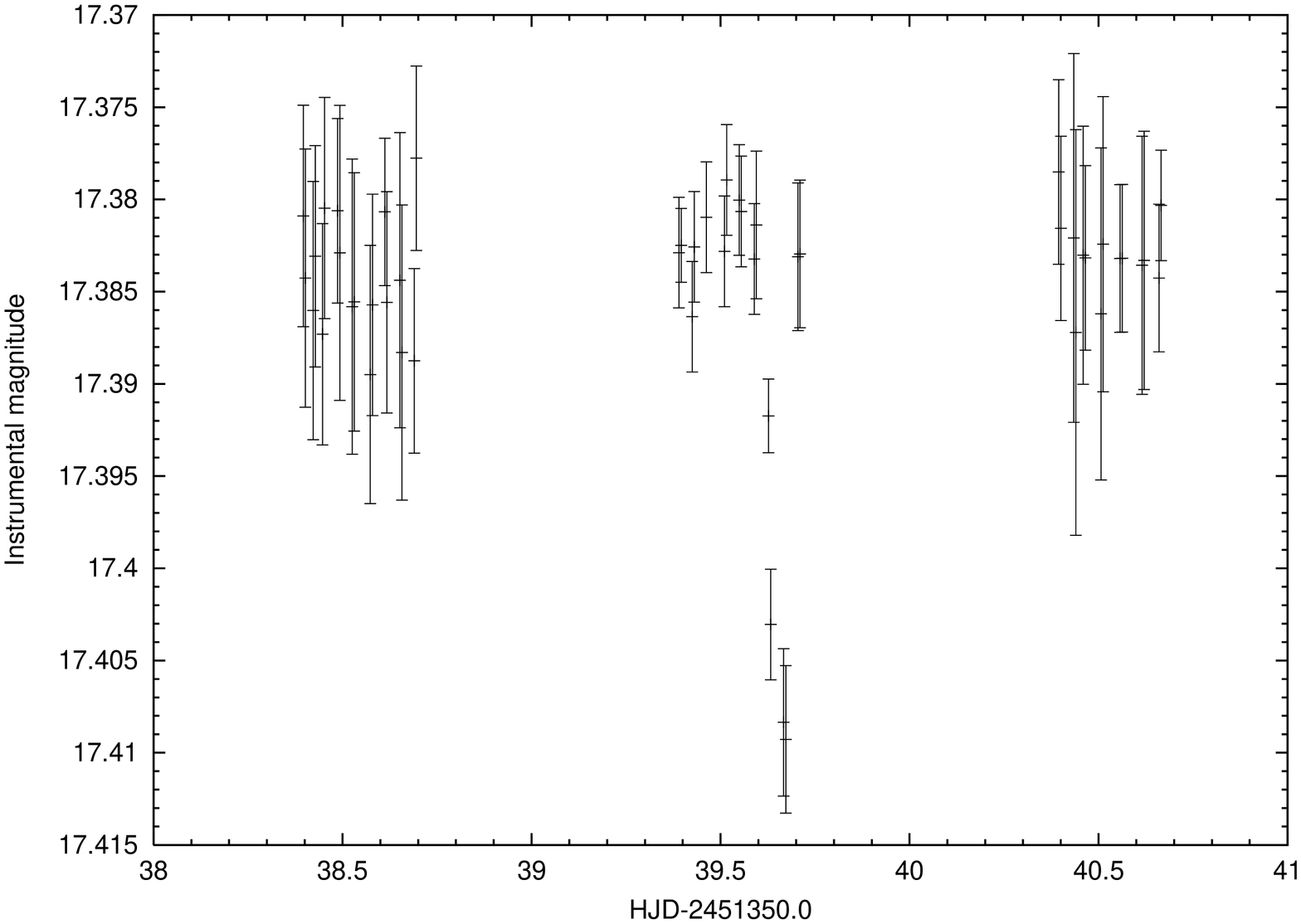}} \\

\end{tabular}
\caption{The lightcurves of stars showing single transit-like eclipse events.  
In the left-hand column, arrows mark the location of the suspected transit.}
\protect\label{fig:singletranslc}
\end{figure*}

Normally we would require at least two transits in a lightcurve in order to
consider a star as a candidate, but the range of possible orbital periods means
that in $\sim$50\% of cases, a transiting planet will only show a single
transit in 20 nights of observations.  The procedure outlined above also
identified 3 lightcurves which appear to show single transit-like eclipses. 
The calculated minimum radii of these companions are all less than 0.3\,\Rsol. 
Table~\ref{tab:singletrans} gives the details of these stars, while
Figure~\ref{fig:singletranslc} displays the full lightcurves next to
lightcurves of the ``transits''.  

\begin{table*}
\centering
\caption{The parameters of the stars which show single transit-like eclipses.}
\protect\label{tab:singletrans}
\vspace{5mm}
\begin{tabular}{cccccccccccc}
\hline
Star	  & $V$    & $V-R$  & $\delta m$  & $\delta t$& $R_{*}$ & $R_{c}$ & Epoch	  &  RA       	& Dec	   \\
	  & (mag)  & (mag)  & (mag)	  & (hours)   & (\Rsol) & (\Rsol) & (HJD-2400000) & (J2000.0)	& (J2000.0)  \\
\hline	
829   	  & 20.178 & 0.500  & 0.04   	  & 2.4       & 1.08  	& 0.22    & 51385.561(9)  & 19 42 07.06 & +39 59 38.9 \\
8153$^a$  & 21.720 & 1.120  & 0.21   	  & 4.8       & 0.63  	& 0.27	  & 51390.522(9)  & 19 40 37.98 & +40 01 04.8 \\
9329$^b$  & 16.743 & 0.534  & 0.03   	  & 2.4       & 1.00  	& 0.17	  & 51389.656(3)  & 19 40 21.71 & +40 04 10.0 \\
\hline
\multicolumn{10}{l}{\parbox{14cm}{\small $^a$ Blended, $^b$ Near saturation.}}
\end{tabular}
\end{table*}

In all three lightcurves the transit-like events could be flat-bottomed,
although better sampled photometry is required to determine this conclusively
and to confirm the events.  The calculated minimum radii suggest brown dwarf
companions; the radius of star 9329 could even be planetary.  We strongly urge
follow-up of these candidates.  

\section{Future Observations} 

\subsection{Follow-up of Transit Candidates}
\protect\label{sec:followup}

Since hot Jupiter planets can have similar radii to brown dwarfs and even
small stars \citet{burrows00}, it is necessary to obtain radial velocity
measurements in order to determine the minimum mass of the companion.  
Together with high precision, continuously sampled lightcurves, the true
companion mass can then be derived.  While this survey has not produced
any clear planetary candidates, several of our companion objects may be
brown dwarfs, and follow-up study would be valuable to confirm or deny
this.  Although the candidates from this survey are much fainter than
those covered by the radial velocity planet hunting surveys, spectroscopic
follow-up of low-radius companions is possible and can provide very useful
information.

Firstly, low resolution spectra of each would provide a more secure spectral
classification (and radius) of the primary than our present estimates based on
broad-band colour indices.  For the faintest candidates ($V\geq18\fmag5$), this
will be the only spectroscopy follow-up possible.  For most of our candidates
however, it is possible to obtain radial velocity measurements using 8--10m
class telescopes (see for example, \citet{yee02}).  While they will not be
precise enough to measure a planetary mass, they will place useful limits on
the mass, confirming or ruling out stellar companions.  

Continuously sampled photometric follow-up is highly desirable, in two colours
if possible.  In our original observation strategy we decided to cycle around
three separate cluster fields in order to cover as many stars as possible.  In
retrospect, we find that continuous observations of a single field is
preferable, in order to get clear, well defined eclipses.   Highly sampled
photometry would clearly reveal the eclipse morphology, distinguish total
eclipses from grazing incidence events and allow detailed models to be fitted. 
For the fainter candidates with no radial velocity observations, this will be
crucial in determining the nature of the system.  Photometry would also improve
the ephemeris, allowing us to time radial velocity observations better.  The
INT/WFC or similarly equiped 2m class telescope could be used for this purpose.

\subsection{Transit Search Strategy}

The results reported here were obtained from our survey's first observing
season, and several improvements to our strategy have now been adopted as a
result.   Firstly, well sampled lightcurves are crucial.  Originally, we tried
to include as many stars as possible by covering 3 clusters in rotation,
resulting in a sampling rate of about 2 observations/star/hour.  A planetary
transit, of typical duration $\sim$2.5\,hours would be represented by perhaps
4--6 datapoints.  We have shown that our algorithm can detect such transits
(see Figure~\ref{fig:transitid}).  However, in practise a greater
signal-to-noise is very desirable.  It also helps in distinguishing stellar
eclipses from transits, in determining the properties of the system, and not
least in calculating an accurate ephemeris for follow-up.  We have now begun
continuously sampled observations.  

From the point of view of detecting transits, blended stars in crowded fields
represent the most significant problem.  The additional scattering caused in
the lightcurve can resemble a transit sufficiently well to distract the
algorithm, and even visual inspection.  Better sampled data will help to
alleviate this, and we are investigating image-subtraction techniques which
should deal with blending more effectively (see for example,
\citet{mochejska02}).   The other major source of false detections is stellar
activity and eclipsing binaries, which obviously the algorithm is very good at
finding.  We are currently investigating improvements which will reject these
stars automatically.  

Another issue raised by this work was cluster membership.  Firstly, it is
difficult to know whether any given star is a member or not.  Although this can
be decided by astrometry, few clusters have been studied in this way, and
usually not to faint enough magnitudes.  The best photometric solution is to
obtain good quality colour data for colour-magnitude and if possible
colour-colour diagrams.  These together with separation from cluster centre
measurements can be used to assign membership probabilities.  Secondly, a
transit survey needs to cover large numbers of stars in its chosen population. 
This survey found that only $\sim$6\% of the stars measured were cluster
members from their colours, amounting to just over 2113 stars out of 38,118. 
This total could be improved slightly by selecting larger radius clusters which
would better cover the field of view and reducing the number of unmeasured
stars due to blending/crowding.  A combination of short/long exposures would
cover stars over a larger range of magnitudes, although some caution is
required not simply to increase the number of unsuitable early-type stars in
the sample.  Ultimately, however, open clusters only contain a few thousand
stars at most.  This highlights the need to survey a number of clusters, in
order to observe enough cluster stars with similar ages and metallicities to be
able to make definitive statements about the planetary population.  The
ultimate aim is then to extend the survey to include a significant number of
clusters covering ranges of age and metallicity, which will reveal the
dependence of planetary formation and evolution on these parameters.  As each
cluster requires around 20 nights of observing time on a 2--4m class telescope,
these aims are best achieved by collaborative efforts between survey teams in
order to obtain sufficient telescope time or else a large dedicated
telescope.  

\section{Conclusions} 
\protect\label{sec:conc} 

We have obtained high-precision photometry on over 38,000 stars in the field of
open cluster NGC~6819.  We have developed an algorithm which can effectively
identify transit-like events in sparsely-sampled data.  This has produced 8
candidates showing multiple transit-like events plus a further 3 candidates
showing single eclipses.  Closer analyses of these lightcurves indicates some of
these candidates could be brown dwarfs, while one has a minimum radius similar
to that of HD~209458b.  Follow-up observations of these candidates are well
worth exploring, especially for the single-transit candidates, as mass limits
could be derived for most of them, allowing us to distinguish their real
nature.  This is particularly important as the periods of these objects are all
5\,days or less.  If brown dwarfs are confirmed among the sample, then they
would fall into the so-called ``brown dwarf desert''.  This in turn might
indicate that the low-mass object population in this field differs from that of
the solar neighbourhood.  

Such a result would be interesting given the lack of transiting planets (and
brown dwarfs) found in the old, metal-poor globular cluster, 47 Tuc. 
\citet{brown01} concluded that the absence of planets might be explained by the
low metallicity, and/or crowded environment serving to disrupt planetary
formation and evolution.  NGC~6819 is comparatively metal-rich and provides a
different environment in which to study the importance of these factors.  Our
rough estimate suggests that we should have detected about 11 transiting
planets in these data, if hot Jupiters are as common as they are in the solar
neighbourhood.  Of course, the transit method favours stars of later spectral
type than the RV technique, so it is possible that planetary frequency
decreases for later spectral type.  We will discuss the significance of this
result in detail in an forthcoming paper.  

\section{Acknowledgements} 

We would like to thank Jasonjot Kalirai for kindly agreeing to share his CFHT
results with us prior to their public release.  This research made use of the
{\sc simbad} database operated at CDS, Strasbourg, France and the {\sc webda}
database operated at University of Lausanne, Switzerland.  RAS was funded by
a PPARC research studentship during the course of this work.  The data
reduction and analysis was carried out at the St.~Andrews node of the PPARC
Starlink project. 

\bibliographystyle{mn2e} 
\bibliography{iau_journals,references}

\end{document}